%% file: ms_astroph.tex
\documentclass[11pt,preprint]{aastex}
\usepackage{graphicx}

\begin{document}

\title{ THE MILLENNIUM ARECIBO 21-CM ABSORPTION LINE SURVEY. II.
PROPERTIES OF THE WARM AND COLD NEUTRAL MEDIA }

\author{Carl Heiles}
\affil{Astronomy Department, University of California,
    Berkeley, CA 94720-3411; cheiles@astron.berkeley.edu}

\author{T.H. Troland}
\affil{Department of Physics and Astronomy, University of Kentucky, 
Lexington, KY 40506; troland@pa.uky.edu}

\begin{abstract} We use the Gaussian-fit results of Paper I to
investigate the properties of interstellar HI in the Solar neighborhood.
The Warm and Cold Neutral Media (WNM and CNM) are physically distinct
components. The CNM spin temperature histogram peaks at about 40 K; its
median, weighted by column density, is 70 K. 

	About $60\%$ of all HI is WNM; there is no discernable change in
this fraction at $z=0$. At $z=0$, we derive a volume filling fraction of
about 0.50 for the WNM; this value is very rough. The upper-limit WNM
temperatures determined from line width range upward from $\sim 500$ K;
a minimum of about $48\%$ of the WNM lies in the thermally unstable
region 500 to 5000 K. The WNM is a prominent constituent of the
interstellar medium and its properties depend on many factors, requiring
global models that include all relevant energy sources, of which there
are many.

	We use Principal Components Analysis, together with a form of
least squares fitting that accounts for errors in both the independent
and dependent parameters, to discuss the relationships among the four
CNM Gaussian parameters. The spin temperature $T_s$ and column density
$N(HI)$ are, approximately, the two most important eigenvectors; as
such, they are sufficient, convenient, and physically meaningful primary
parameters for describing CNM clouds.  The Mach number of internal
macroscopic motions for CNM clouds is typically about 3 so that they are
strongly supersonic, but there are wide variations. We discuss the
historical $\tau_0$-$T_s$ relationship in some detail and show that it
has little physical meaning. 

	We discuss CNM morphology using the CNM pressure known from UV
stellar absorption lines. Knowing the pressure allows us to show that
CNM structures cannot be isotropic but instead are sheetlike, with
length-to-thickness aspect ratios ranging up to about 280. We present
large-scale maps of two regions where CNM lies in very large ``blobby
sheets''. 

	We test the McKee/Ostriker model of the interstellar medium by
explicitly modeling our data with CNM cores contained in WNM envelopes.
This modeling scheme works quite well for many sources and also predicts
the WNM filling factor reasonably well. However, it has several
deficiencies. 

\end{abstract}


\tableofcontents

\section{ INTRODUCTION}

\label{introduction}

	This paper discusses  the astronomically oriented results of a
new  Arecibo\footnote{The Arecibo Observatory is part of the National
Astronomy and Ionosphere Center, which is operated by Cornell University
under a cooperative agreement with the National Science Foundation.}
21-cm absorption line survey; it is the comprehensive version of the
preliminary report by Heiles (2001b). Paper I (Heiles and Troland 2002)
discusses the observational and data reduction techniques.

We took great care in accounting for instrumental gain fluctuations and
angular structure of HI so that we could derive accurate opacity and
expected emission profiles, including realistic uncertainties.  (An
expected profile is the emission profile towards the source that would
be observed if the source flux were zero).  The opacity profiles come
from the Cold Neutral Medium (CNM) and are characterized by distinct
peaks; we decomposed them into Gaussian components.  The expected
profiles are produced by both the Warm Neutral Medium (WNM) and the CNM.
 We fit them using a simple but physically correct radiative transfer
equation that includes both the emission and absorption of the CNM and,
in addition, one or a few independent Gaussians for the WNM emission. 
We discussed the fitting process and its uncertainties in detail, and
presented many examples of the technique.  We derived spin temperatures
for the CNM using the opacity and expected profiles.  We derived
upper-limit temperatures for the CNM using the line widths.  We
presented all results in tabular, graphical, and electronic form. 

	Table \ref{sourcelist} summarizes the sources observed and the
column densities of CNM and WNM. Here, by ``WNM'', we mean Gaussian
components detected only in emission, and by ``CNM'' we mean Gaussians
that were detected in absorption.  Paper I presents the full table
of Gaussian component properties. We have a total of 79 sources, 202 CNM
components, and 172 WNM components. 13 sources have $|b|<10^\circ$, and
we exclude these from some of our discussion below because their
profiles are complicated or the WNM linewidths might be significantly
broadened by Galactic rotation. 

	\S \ref{tempdist} shows that the division between WNM and CNM is
not only observational, but also physical; \S \ref{cnmwnmsummary}
summarizes the statistics on CNM/WNM column densities for the Gaussians.
\S \ref{coldenstatistics} presents column density statistics for the
lines of sight for the CNM and WNM. \S \ref{volfill} discusses the
volume filling fraction of the WNM, both at high and low $z$. 

	The next few sections discuss the basic statistical properties
of the Gaussian components. \S \ref{vlsrstatistics} presents statistics
on $V_{LSR}$. \S \ref{correlations} presents correlations among the four
parameters that describe the CNM components. The reader interested in
these correlations should consult the two subsequent sections: \S
\ref{pequality} shows that inadequate angular resolution might affect
these correlations, and \S \ref{againstraisin} shows that CNM features
are sheetlike and not isotropic with the consequence that angular
resolution effects are far less important than found in \S
\ref{pequality}. 

	\S \ref{momodel} re-reduces all the data of Paper I in terms of
the McKee \& Ostriker (1977) (MO) model, with each CNM component
surrounded by an independent WNM component; it is gratifyingly
successful for most sources but some MO predictions are not
quantitatively fulfilled. \S \ref{descmodel} presents two descriptive
models; the second, the clumpy sheet model for the CNM, applies to our
data. 

	\S \ref{summary} is a summary, and \S \ref{commentary} is a
commentary on the importance of the WNM for understanding not only the
ISM but also its multiplicity of energy sources and the Universe at
large.

\section{ THE CNM: AN OBSERVATIONALLY AND PHYSICALLY DISTINCT
TEMPERATURE COMPONENT}

\subsection{ Distribution of CNM and WNM spin and kinetic temperatures
for $|b| > 10^\circ$}

\label{tempdist}

	At $|b|>10^\circ$, for the WNM we have 143 components from 66
lines of sight, each of which is a radio source, containing a total
$N(HI)_{WNM,20}=292$ and for the CNM we have a total of 143 components
from 48 sources containing a total $N(HI)_{CNM,20}= 188$; the subscript
20 on $N(HI)$ means that the units are $10^{20}$ cm$^{-2}$. There
are fewer CNM sources because 18 sources had undetectable absorption.

	For the CNM we have direct, fairly accurate measurements of
$T_s$ derived from the fitting process described in Paper I (\S 4.3). 
For the CNM, the spin temperature is equal to the kinetic temperature. 
For the WNM we have rough {\it lower} limits on $T_s$ from the absence
of WNM absorption in the opacity profiles.  For both the CNM and WNM, we
have {\it upper} limits on kinetic temperature $T_{kmax}$ from the line
width.  For warm, low-density gas $T_s$ is not necessarily equal to the
kinetic temperature, with $T_s < T_k$; for equilibrium conditions, this
inequality becomes serious only for $T_k \gtrsim 1000$ K (Liszt 2001). 
Thus, our lower limit on $T_s$ is also a lower limit on $T_K$, so $T_K$
is bracketed; and for $T_s \lesssim 1000$ K, $T_s \approx T_k$. 

\begin{figure}[h!]
\begin{center}
\includegraphics[width=3.5in] {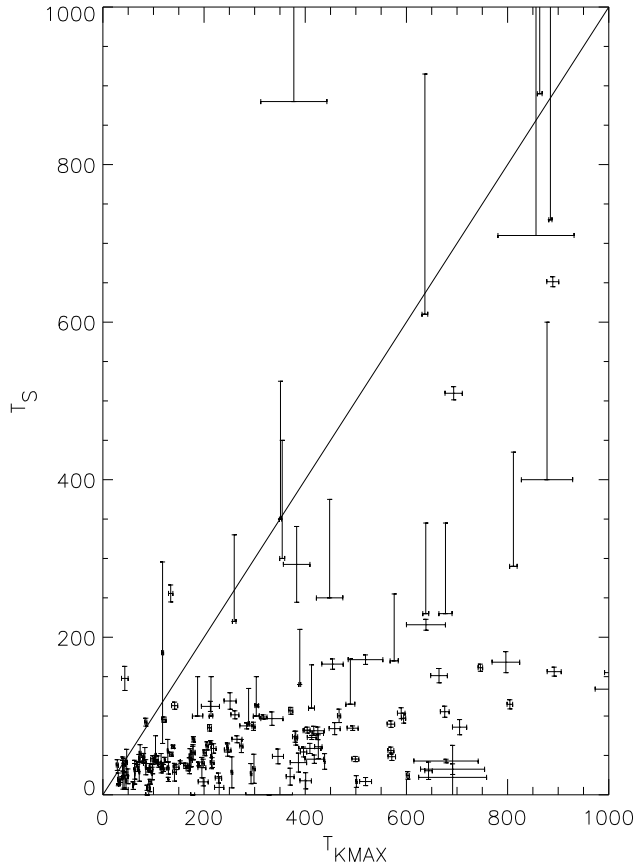} 
\end{center}

\caption{ Spin temperature $T_s$ versus upper-limit kinetic temperature
$T_{kmax}$ for all Gaussian components, both CNM and WNM, for sources
having $|b| > 10^\circ$. WNM errorbars only go up because they are lower
limits. \label{tkintspin}}  \end{figure}

	Figure \ref{tkintspin} compares either $T_s$ (CNM components) or
lower limits on $T_s$ (WNM components) with $T_{kmax}$ for every
Gaussian component at $|b|>10^\circ$.  For the CNM components we show
errorbars for $T_s$; for the WNM component $T_s$ is a lower limit, so
its errorbars go in only one direction and are arbitrarily set to be
half the estimated value.  Because $T_s \leq T_k$ and $T_{kmax} \geq
T_k$, the points should all fall below the diagonal line.  Nearly all of
them do.  There are five serious exceptions for which the difference is
significantly larger than the error: a CNM component in each of 3C123,
3C237, and 4C32.44; and a WNM component in each of 3C93.1 and NRAO140. 
The profiles of all these sources are complicated, increasing the chance
that the choice of Gaussians is not realistic.  Thus there is general
agreement with the requirement that all points fall below the line.  In
fact, most points fall well below the line, particularly for the CNM. 

%\clearpage

%\clearpage
\begin{figure}[p!]
\begin{center}
\includegraphics[width=5in] {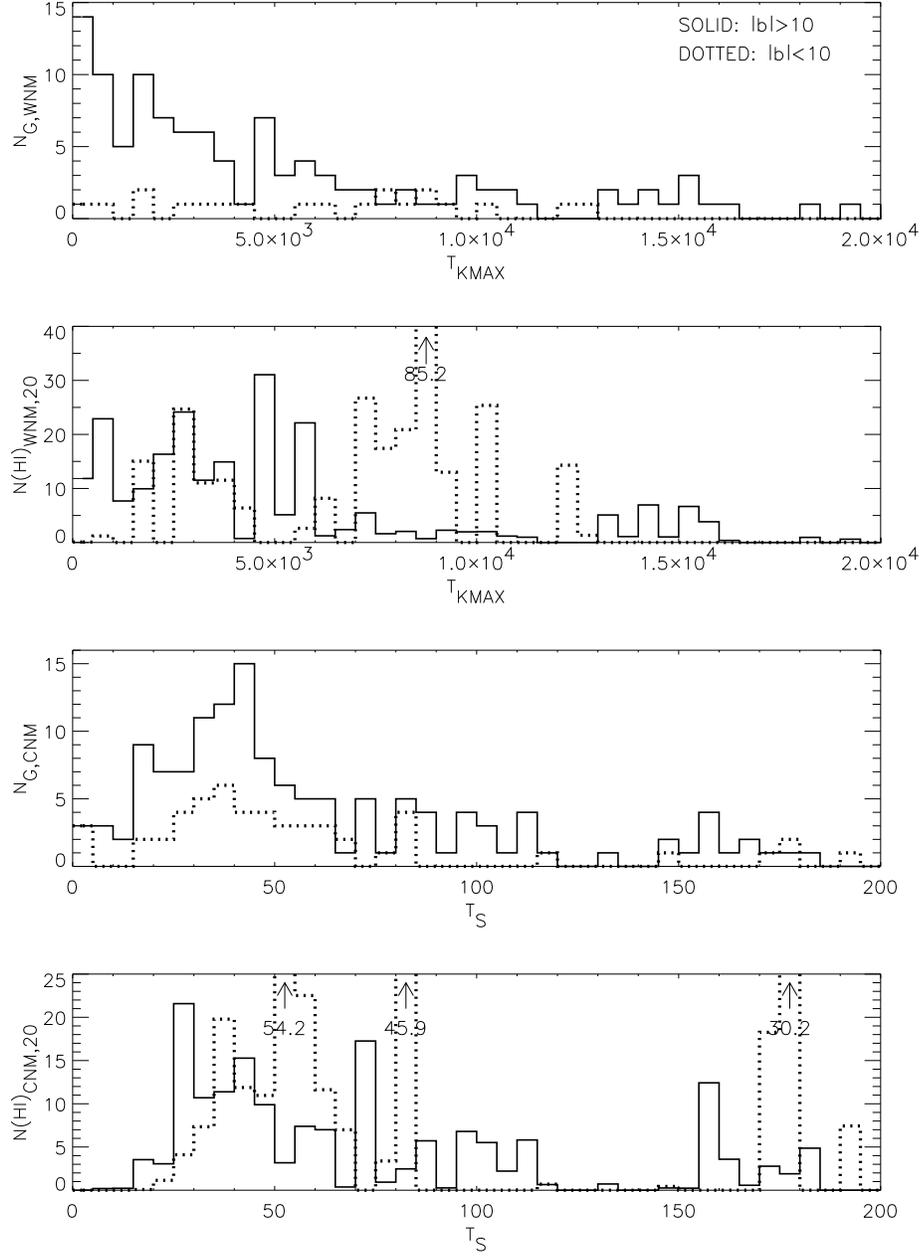} 
\end{center}

\caption{Histograms of $T_{kmax}$ for the WNM (top two panels), and of
$T_s$ for the CNM (bottom two panels). The solid lines are for
$|b|>10^\circ$ and the dotted ones for $|b|<10^\circ$. $N_G$ is number
of Gaussian components and $N(HI)_{20}$ is column density in units of
$10^{20}$ cm$^{-2}$. \label{histoplot4new}}  \end{figure}
%\clearpage

	Figure \ref{histoplot4new} displays the temperature
distributions of the WNM and the CNM.  The upper two plots are for the
WNM where we plot both the number of Gaussian components $N_{G,WNM}$,
and the column density of these Gaussians $N(HI)_{WNM,20}$, versus
$T_{kmax}$.  The two lower plots show the analogous temperature
distributions for the CNM, plotted versus $T_s$.  In all cases, solid
and dotted lines are for $|b|>10^\circ$ and $|b|<10^\circ$,
respectively.  We separate the Galactic plane sources having
$|b|<10^\circ$ for three reasons: (1) their profiles have high column
densities and dominate the $N(HI)_{20}$ histograms; (2) their profiles
are broadened by Galactic rotation, unphysically increasing $T_{kmax}$;
(3) their spin and upper-limit kinetic temperatures can be distorted by
uncertainties in the fits because the profiles are sometimes so
complicated. 

	For the WNM in the top two panels, a significant fraction of the
WNM gas has $500 < T_{kmax} < 5000$ K, which puts it in the thermally
unstable range.  $N_{G,WNM}=14$ WNM components and $N(HI)_{WNM,20}=11.8$
have $T_{kmax} < 500$ K, so can be classed as too cold to be thermally
unstable; these correspond to ($N_{G,WNM}, N(HI)_{WNM,20})$ fractions
($10\%, 4\%)$, respectively.  The unstable range has fractions $(39\%,
48\%)$.  Even though the lower limits on $T_s$ for some of this gas lie
below 500 K, we regard as very remote the possibility that $T_s$ is
actually so low because it would require highly supersonic motions. 
Under this assumption, this is the fraction of WNM gas that truly lies
in the unstable range.  Most of the rest $(28\%, 26\%)$ lies between
5000 and 20000 K, and $(23\%, 22\%)$ have $T_{kmax} > 20000$ K and lie
off the histograms shown.  Any gas having $T_k \gtrsim 10000$ K would be
ionized, so components having $T_{kmax} > 10000$ K must either consist
of multiple blended narrower components or must have highly supersonic
motions. 

	For the CNM in the bottom two panels, the histograms exhibit
well-defined broad peaks near 40 K. Most of the gas $(77\%, 67\%)$
has $T_s < 100$ K.  Some of the gas $(17\%, 4\%)$ is very cold, with
$T_s < 25$ K; this cannot occur unless photoelectric heating by dust is
inoperative (Wolfire et al 1995).  In these histograms, the fractions
having $T_s > 200$ K and lying off of the histogram are $(8\%,
11\%)$, with the maximum $T_s=656$ K. 

\subsection{ CNM and WNM combined: distinct populations}

	Here we address the question of whether the  CNM comprises a
distinct temperature population. Of course, the CNM is observationally
distinguished by its detection in opacity profiles; however, this
depends on sensitivity and does not necessarily mean that its belongs to
a distinct physical population in the ISM. We restrict our attention to
sources having $|b| > 10^\circ$ to minimize the artificial increase of
$T_{kmax}$ caused by Galactic rotation and to reduce uncertainties from
incorrectly-modeled blended components. 

%\clearpage
\begin{figure}[p!]
\begin{center}
\includegraphics[width=5in] {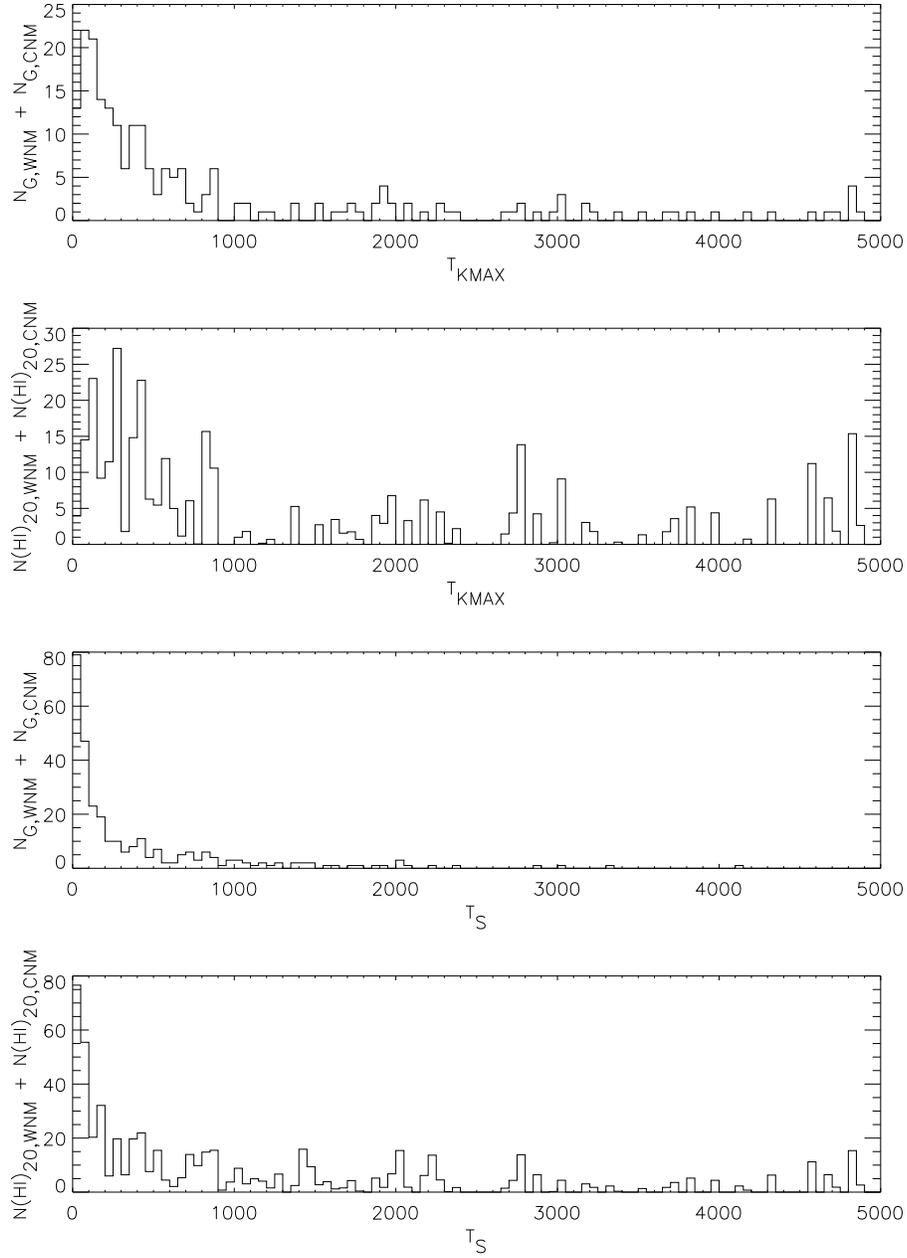} 
\end{center}

\caption{ Histograms of upper-limit kinetic temperatures $T_{kmax}$ and
spin temperatures $T_s$ for the combined set of WNM and CNM components
for sources having $b|>10^\circ$. $N_G$ is the number of Gaussian
components; $N(HI)_{20}$ is the column density in units of $10^{20}$
cm$^{-2}$. For WNM components, spin temperatures are lower limits.
\label{histoplot7}}  \end{figure}
%\clearpage

	Figure \ref{histoplot7} lumps all temperatures, both CNM and
WNM, into a single distribution and provides histograms for both number
of Gaussian components $N_G$ and column density $N(HI)_{20}$. First consider
the first (top) and third panels, which are the histograms of $T_{kmax}$
and $T_s$ for $N_G$. Both panels exhibit a strong peak towards the left
and a long, flat distribution towards the right. These shapes are not
suggestive of a continuous distribution, but rather two distributions:
one peaked at low temperatures and one spread roughly uniformly over a
very broad temperature range running well above 5000 K.
The low-temperature peak in $T_s$ for $T_s \lesssim 200$ K is nearly
all CNM components; the highest CNM temperature is 656 K. Similar
comments apply to panels two and four, which are the histograms for
$N(HI)_{20}$, but these histograms are noisier.

	We conclude that the CNM is indeed a separate, distinct
temperature distribution in the ISM. The median temperature for its
Gaussian components is 48 K and for column density 70 K (Table
\ref{medianmeants}), but the histogram in Figure \ref{histoplot4new}
shows large variations. The physical division between the two ISM
temperature components is operationally the same as the division between
CNM and WNM. However, CNM components lying at high temperatures could
also be considered as very cool WNM; the boundary is a bit blurred.

\subsection{Column density statistics for WNM and CNM Gaussian components}

\subsubsection{ Histograms}

%\clearpage
\begin{figure}[p!]
\begin{center}
\includegraphics[width=5in] {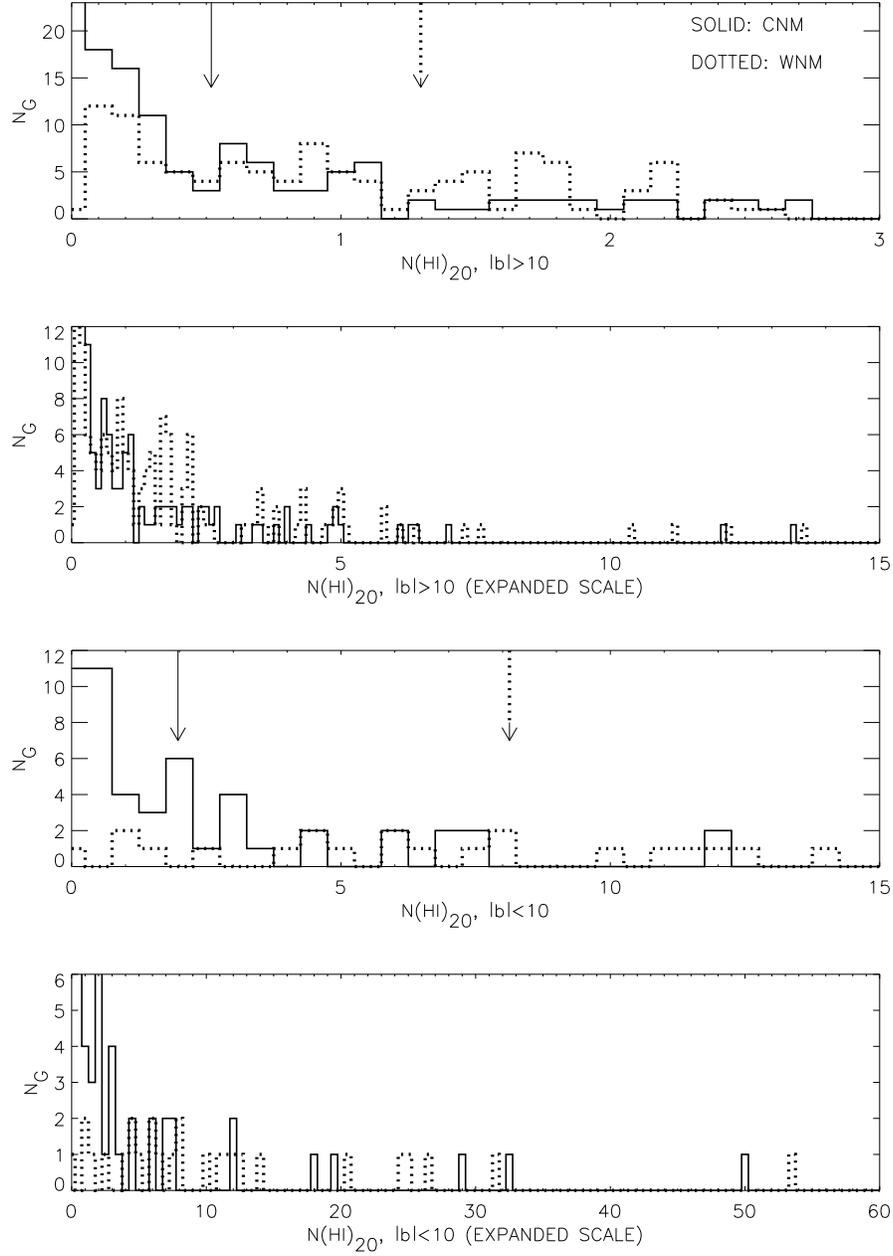} 
\end{center}

\caption{Histograms of number of Gaussians $N_G$ and column densities
$N(HI)_{20}$ for all Gaussian components, both CNM (solid) and WNM
(dotted). The top two panels show $|b|>10^\circ$ with different scales
on both axes to facilitate interpretation; the bottom two panels are for
$|b|<10^\circ$. The arrows show the medians, which are for the (CNM,
WNM) (0.60, 1.30) at $|b|>10^\circ$ and (2.0, 5.0) at  $|b|<10^\circ$.
N$(HI)_{20}$ is in units of $10^{20}$ cm$^{-2}$. \label{histoplotnhgaussians}}
\end{figure}
%\clearpage

	Figure \ref{histoplotnhgaussians} exhibits separate histograms
of $N(HI)_{20}$ for the CNM and WNM Gaussian components.  The top two
panels show $|b|>10^\circ$ with different scales on both axes to
facilitate interpretation; the bottom two panels are for $|b|<10^\circ$.
 Table \ref{medianmeannh} gives the medians and means. The ranges of
column density are enormous, covering more than a factor 100.  At low
latitudes we see many fewer Gaussian components having $N(HI)_{CNM,20}
\lesssim 0.5$, possibly because they are indistinguishable in the
presence of blended components at low latitudes. 

	There appears to be an excess or independent population of
low-column-density CNM components having $N(HI)_{CNM,20} < 0.5$;
otherwise, CNM and WNM components have similar column density
distributions at both high and low latitudes. The similarity of the
WNM and CNM distributions for $N(HI)_{CNM,20} > 0.5$ suggests that 
the two phases could be part of the same population and that members can
adopt either temperature range according to circumstances. 

\subsubsection{Overall summary statistics}

\label{cnmwnmsummary}

	For sources at $|b|>10^\circ$, the global ratio of WNM to total
HI column density is $\langle R(HI)_{WNM} \rangle =0.61$.  Mass is
equivalent to column density if the distances are the same.  The WNM is
systematically more distant than the CNM because it has a larger scale
height (Kulkarni \& Heiles 1987), so this is a lower limit for the mass
fraction. 

	The $N(HI)$ fraction of WNM having $T_{kmax}$ in the unstable
region 500 to 5000 K is $0.48$; the true fraction of gas in this
unstable regime might be higher because $T_{kmax}$ is an upper limit on
temperature derived from the linewidth.  It's conceivable, but unlikely
in our opinion, that much of this gas has temperature $T_k<500$ K.  The
$N(HI)$ fraction of CNM having $T_s$ in the range 25 to 70 K (the main
peak in the histogram) is $0.46$. 

	At low latitudes, $|b| \lesssim 10^\circ$, the line of sight
doesn't leave the HI layer for nearby gas. We can use low-latitude
sources as a test to determine whether the fraction of WNM gas
$R(HI)_{WNM}$ decreases at lower $|z|$ where the pressure is higher, as
is theoretically predicted. We have 8 sources with reasonably accurate
Gaussian fits (and 5 with unusable fits; Table \ref{sourcelist}). These
8 sources have $\langle R(HI)_{WNM}\rangle = (0.67 \pm 0.08)$. This is
indistinguishable from the $|b|>10^\circ$ mean value $\langle
R(HI)_{WNM}\rangle = 0.61$. Thus, there is no evidence for the predicted
decrease in $R(HI)_{WNM}$. However, we stress that our low-latitude
results are generally less accurate than the others because it is more
difficult to obtain accurate expected profiles and to perform Gaussian
fits. Accurate results for low latitudes probably requires
high-sensitivity interferometric observations.

\subsection{Comparison of CNM temperatures with other results}

\label{comparison}

\subsubsection{ Previous CNM temperatures from the 21-cm line}

	Our spin temperatures are colder than previously obtained ones.
Histograms of CNM temperatures have been given by Dickey, Salpeter, \&
Terzian (1978, DST), Payne, Salpeter, \& Terzian (1983, PST), and Mebold
\ et al 1982, among others. They find broader histograms than ours with
temperatures extending to much higher values and median values in the
neighborhood of 80 K; for example, Mebold et al find a median (by
components) of 86 K. Our histogram is narrower and peaked near 40 K
(Figure \ref{histoplot4new}) and our median (by components) is 48 K. In
contrast, our median (weighted by column density) is 70 K. When quoting
medians, it is important to distinguish between the component median and
the column-density median. 

	Our lower temperatures do not arise because the older data were
incorrect (although some were); it is because the analyses were
incorrect. In contrast to the previous treatments, our Gaussian
technique, which is thoroughly discussed in Paper I \S 4 and \S 5,
properly accounts for the two-phase medium and the associated radiative
transfer. Recent measurements of temperatures in the Magellanic Clouds
(Mebold et al 1997; Marx-Zimmer et al 2000; Dickey et al 2000) use the
slope technique, which also properly treats radiative transfer for
simple profiles (Paper I, \S 4, \S 6); they find smaller temperatures,
consistent with ours, and show that the older incorrect technique yields
incorrect higher temperatures.

\subsubsection{ Temperatures from H$_2$}

	Temperatures are also derived from the ratio of populations in
the two lowest rotational states of H$_2$.  Unfortunately, these are not
directly comparable to our CNM temperatures, for two reasons. First, the
H$_2$ lines of sight are chosen to maximize column density; in contrast,
ours are random with respect to column density. Second, the H$_2$ lines
are saturated, which means that the derived temperatures are a weighted
average over all velocity components and all the gas, both CNM and WNM;
one cannot know which phase dominates the results because the fractional
H$_2$ abundances in the two phases are unknown. Because the H$_2$
measurements refer to all gas, a median derived therefrom is more akin
to a column-density median than a component median.

	Recent FUSE measurements (Shull et al 2000) confirm the large
survey of Savage et al (1977), who found the range of temperatures to be
$T_{H_2} = 77 \pm 17$ (rms) K.   This is comparable to our component
median for the CNM. However, because the H$_2$ sample is biased to large
column density lines of sight, the results are not directly comparable.
We further explore the comparison by considering four of our sources
that are fairly close to stars in three regions studied by Savage et al.
 This by no means guarantees that the physical regions sampled are
identical, but one hopes that the lines of sight are physically
similar.Table \ref{h2tbl} shows radio sources and stars in these three
areas; in each area the radio and optical positions are close, within a
few degrees.  The first two regions have high $N(HI)$ and are cold, with
CNM temperatures lying near the peak of our histogram; the H$_2$
temperatures are higher than the HI temperatures.  We detected the 21-cm
line in absorption in the third region but we would not classify the
$510$ K gas as CNM; the H$_2$ temperature of 377 K is smaller than the
HI temperature, although realistic uncertainties may mean that the
results are consistent. 

	The upshot is that the H$_2$ temperatures do not agree with the
HI CNM temperatures: This conclusion needs confirmation via
observations of HI and H$_2$ absorption along identical lines of sight.
Such observations require a background source such as 3C273 with
significant radio and UV emission.

\section{STATISTICS ON INTEGRATED LINE-OF-SIGHT HI COLUMN DENSITY}

\label{coldenstatistics}

%\clearpage
\begin{figure}[h!]
\begin{center}
\includegraphics[width=3.5in] {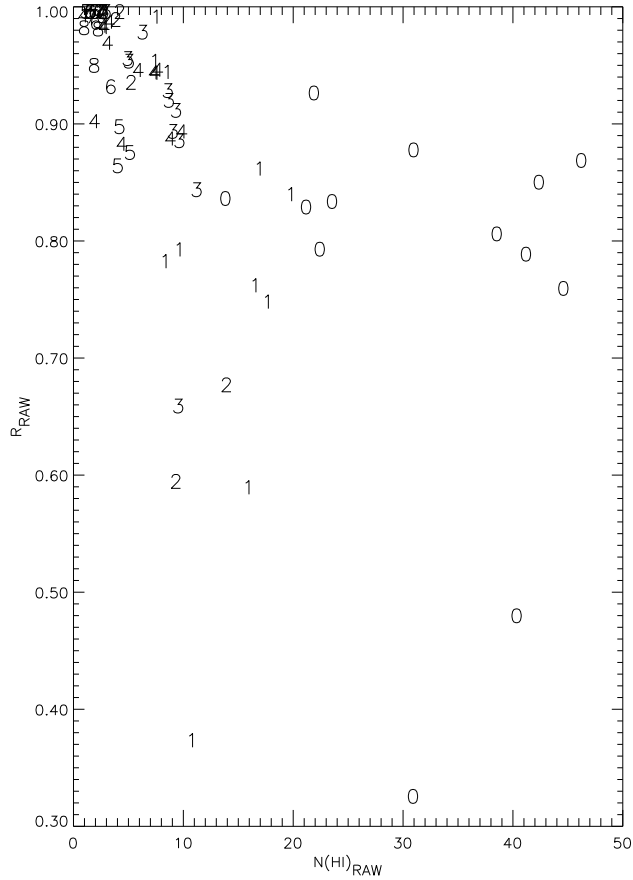} 
\end{center}

\caption{Plot of the ratio $R_{raw}= N(HI)_{raw}/N(HI)_{tot}$ versus
$N(HI)_{raw}$ for our lines of sight; units are $10^{20}$ cm$^{-2}$. 
This is the factor by which HI column densities obtained from brightness
profile integrals are too small. Numbers are $int(|b/10|)$; for example,
3 means $|b|$ lies between $30^\circ$ and $40^\circ$. 
\label{mapnhrawratio2}} \end{figure}

%\clearpage

\begin{figure}[p!]
\begin{center}
\includegraphics[width=5in] {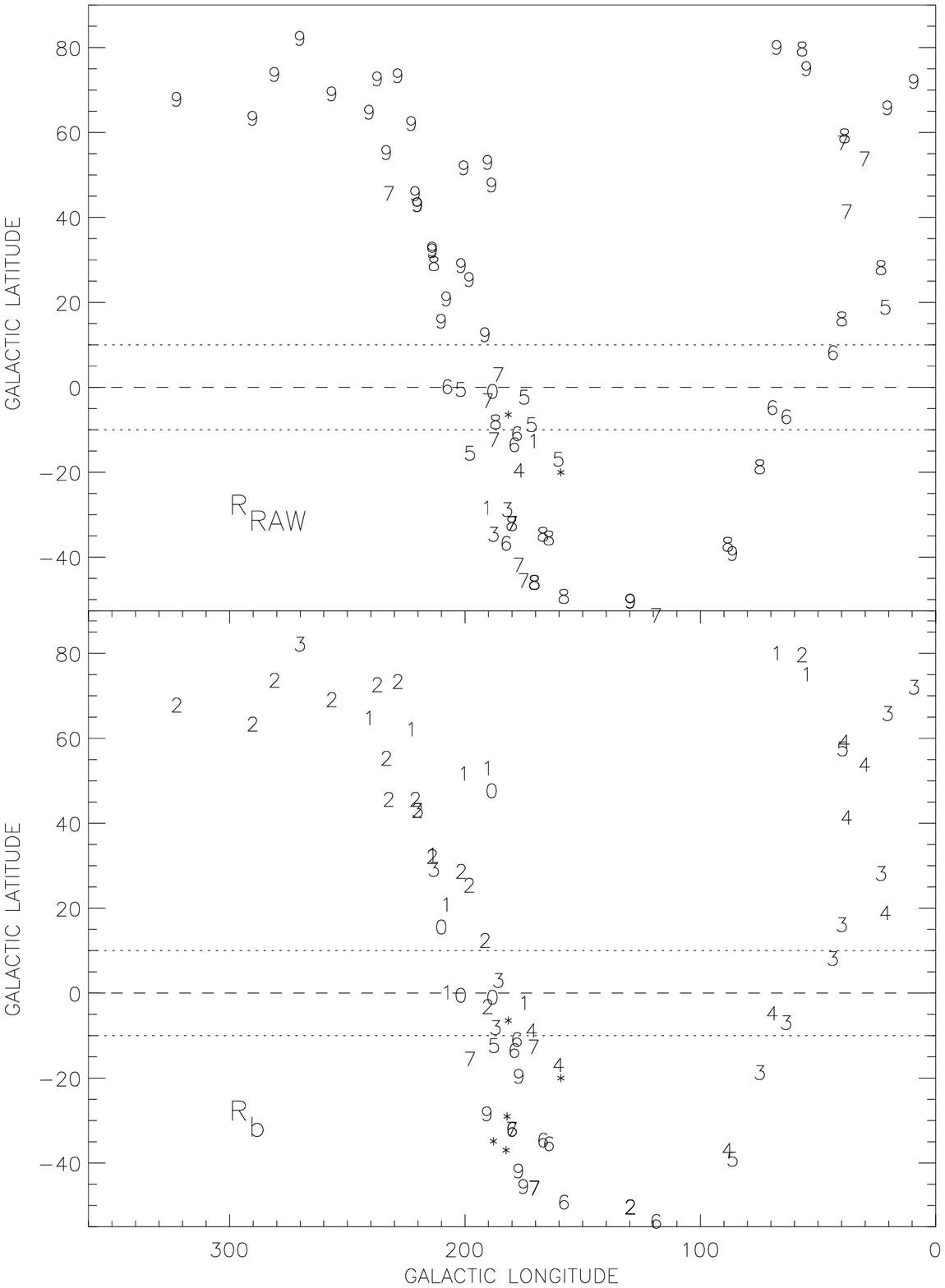} 
\end{center}

\caption{{\it Top}: Map of the ratio $R_{raw}= N(HI)_{raw}/N(HI)_{tot}$
for our lines of sight. Numbers are $int(20 (R_{raw}-0.5)$; for example,
7 means $R_{raw}=0.85$ to 0.9. {\it Bottom}: Map of $int(4.5 R_b)$, 4.5
times the ratio of actual to total column density expected for a smooth
plane-parallel layer in the Galaxy (equation \ref{planeparallel}). For
example, 4 means $R_b= 0.90$ to 1.11. Asterisks mean numbers exceed 9.
\label{mapnhrawratio} \label{mapbeeratio}} \end{figure}
%\clearpage

\subsection{Raw versus true HI column density}

	One is often interested in the total HI column density. One
calculates this from 21-cm line data by assuming that $\tau(\nu) \ll 1$;
then $N(HI) \propto$ the profile area. An accurate calculation for the
general case requires knowledge of the opacity and the arrangement of
the absorbing clouds along the line of sight, which our analysis
technique provides. We use our results to compare these two methods.

	We define the ``raw'' HI column density $N(HI)_{raw}$ as that
obtained from the profile area. The true HI column density for a line of
sight is equal to  $N(HI)_{tot}= \sum N(HI)_{CNM} + \sum N(HI)_{WNM}$,
where $\sum$ means summed over all Gaussian components for a line of
sight. The ratio 

\begin{equation}
R_{raw}= N(HI)_{raw}/N(HI)_{tot}
\end{equation}

\noindent is plotted versus $N(HI)_{raw}$ in Figure
\ref{mapnhrawratio2}.  Numbers indicate the Galactic latitude $|b|$ in
units of 10 degrees.  Significant corrections exist, in some cases even
at high latitudes and low measured column densities. 

	Figure \ref{mapnhrawratio} (top) shows a map of $R_{raw}$ in
which the numbers are $int[20( R_{raw}-0.5)]$; for example, 7 means
$R_{raw}=0.85$ to 0.9.  Areas of sky are characterized by $R_{raw}$. 
For example, the Taurus/Perseus region ($l=155^\circ$ to $180^\circ$,
$b=-25^\circ$ to $-10^\circ$) has uniformly small values, which is not
surprising because of the many molecular clouds and overall high column
densities. 

\subsection{Statistics on line-of-sight HI column densities for
$|b|>10^\circ$}

	The effect of local structures on total column density is much
stronger than the expected latitude dependence.  This prevents us from
analyzing column density statistics in the usual way of accounting for
the expected latitude dependence.  In our plane-parallel Galaxy, one
classically expects the total column density to be $N(HI)_{20} =
3.7/\sin |b|$ (Kulkarni \& Heiles 1987).  Define the ratio of the true
measured column density to this expected value

\begin{equation} \label{planeparallel}
R_b = { N(HI)_{tot,20} \over 3.7/\sin(|b|) }
\end{equation}

\noindent Figure \ref{mapbeeratio} (bottom) is a map of $int(4.5 R_b)$;
for example, a number 4 means $R_b=0.89$ to 1.11, so all of the numbers
on this map should be equal to 4.  Clearly, some areas of sky are
deficient and some overabundant.

%\clearpage
\begin{figure}[p!]
\begin{center}
\includegraphics[width=5in] {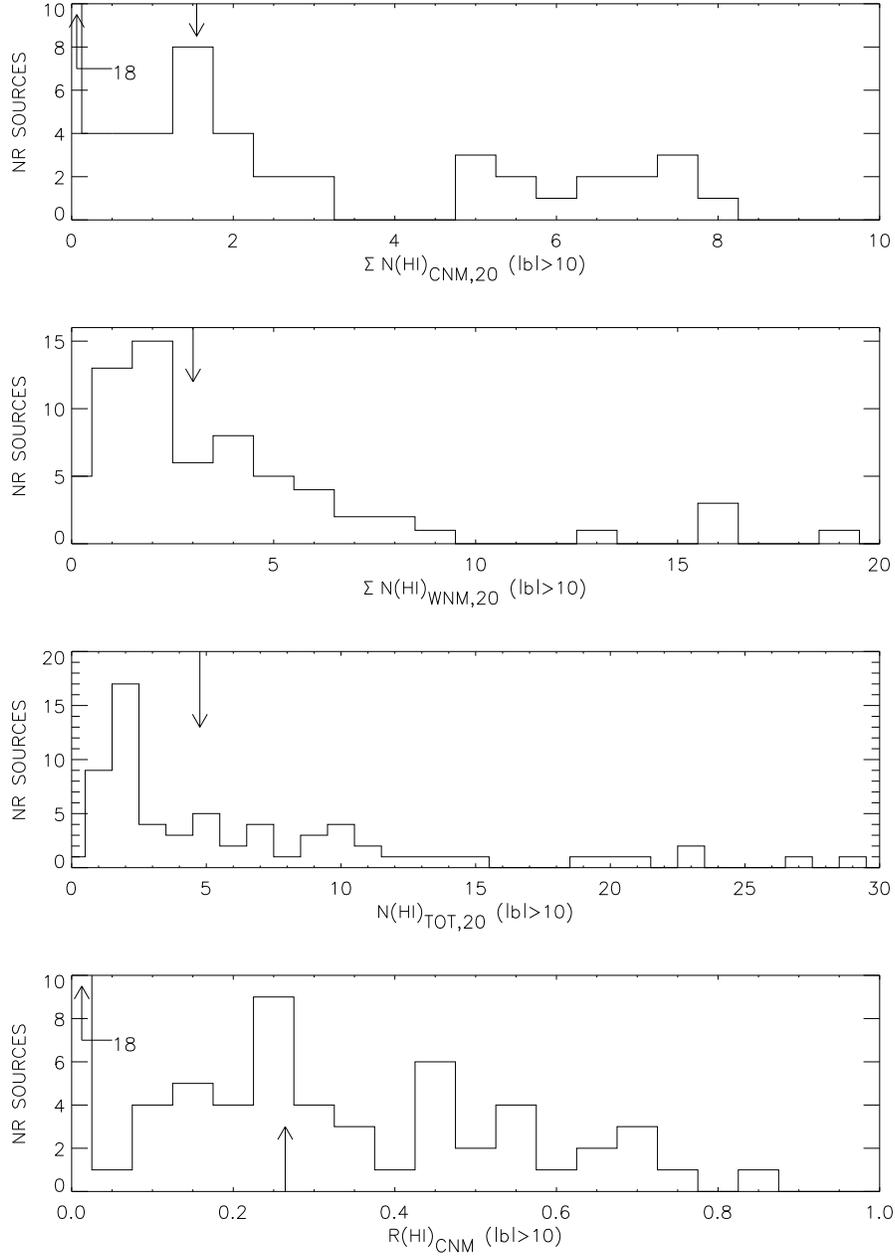} 
\end{center}

\caption{Histograms of $\sum N(HI)_{CNM,20}$ and $\sum N(HI)_{WNM,20}$,
the total column densities for each line of sight, for sources
having $|b|>10^\circ$. We plot the CNM and WNM individually (top two
panels); the total $N(HI)_{tot,20}= \sum N(HI)_{CNM,20} + \sum
N(HI)_{WNM,20}$ (third panel), and the CNM  fraction $R(HI)_{CNM} = (\sum
N(HI)_{CNM})/N(HI)_{tot}$.  Arrows show the medians. \label{histnh}}
\end{figure}
%\clearpage

	The top two panels of Figure \ref{histnh} exhibit the histograms
of $\sum N(HI)_{CNM,20}$ and $\sum N(HI)_{WNM,20}$  individually; within
the statistics the shapes are not too dissimilar, but the WNM column
densities are about twice the CNM ones. The third panel exhibits the
histogram for $N(HI)_{tot,20}$; the low-$N(HI)_{tot}$ peak is from the
CNM and the tail from the WNM. The fourth panel exhibits the histogram
of the CNM column density fraction 

\begin{equation}
R(HI)_{CNM}= { \sum N(HI)_{CNM} \over N(HI)_{tot} }
\end{equation}

\noindent for each line of sight. 

	The fourth panel, together with the top panel, show a huge peak
with zero  $\sum N(HI)_{CNM}$. In each case, the peak is distinct from
the rest of the histogram.   Therefore, lines of sight having zero $\sum
N(HI)_{CNM}$ form a {\it distinct class}. Lines of sight to the majority
of sources have $R(HI)_{CNM} \leq 0.3$; however, a few lines of sight
are dominated by CNM.  

\begin{figure}[h!]
\begin{center}
\includegraphics[width=3.5in] {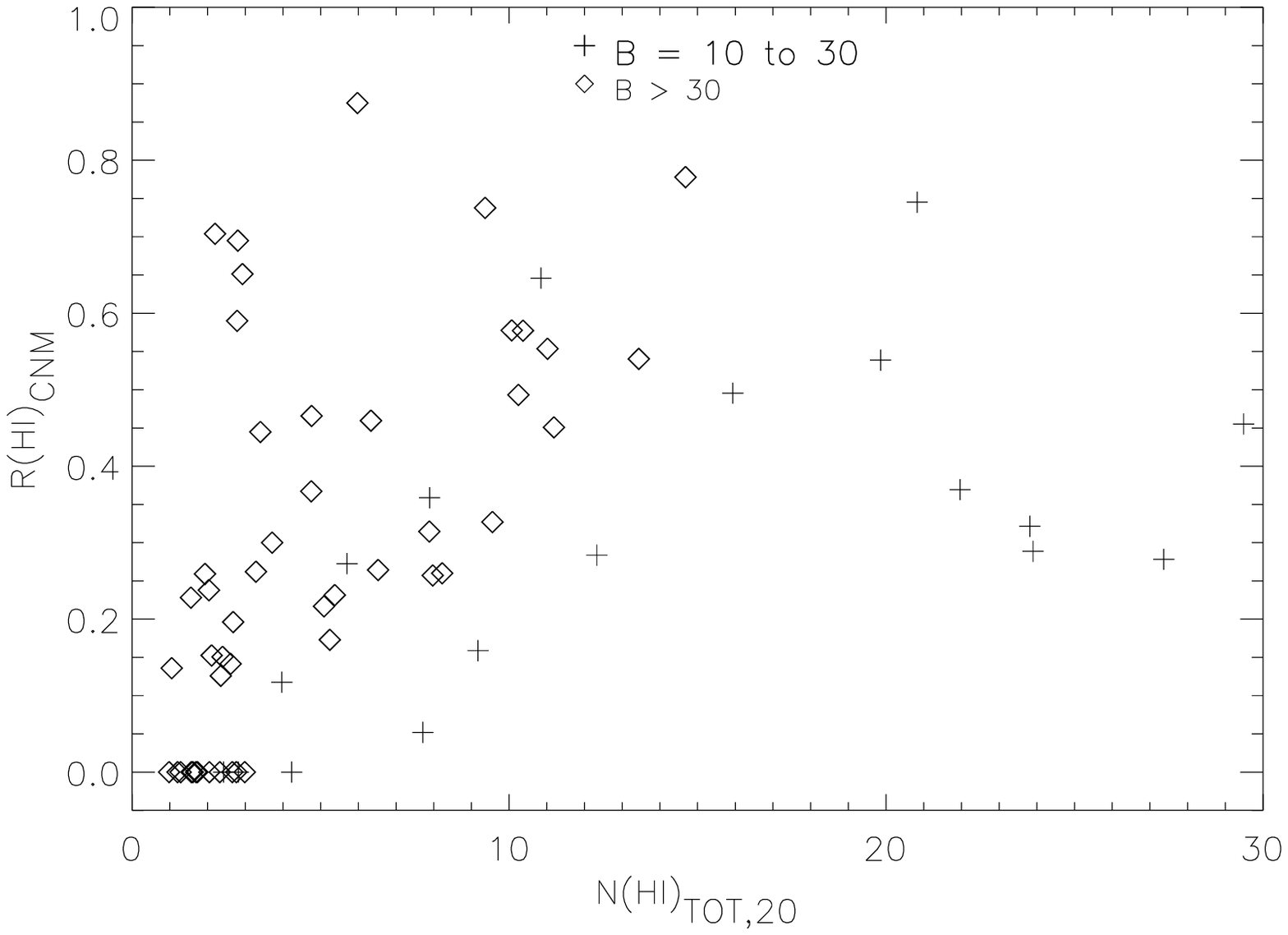} 
\end{center}

\caption{The CNM fraction $R(HI)_{CNM}=N(HI)_{CNM}/ (N(HI)_{tot}$ versus
$N(HI)_{tot,20}$, with low latitudes differentiated from high ones by the
diamond and plus-sign symbols.  For a map of $R(HI)_{CNM}$, see Figure
\ref{mapnhratio}. \label{nhtotalnhratio}} 
\end{figure}

\begin{figure}[p!]
\begin{center}
\includegraphics[width=5in] {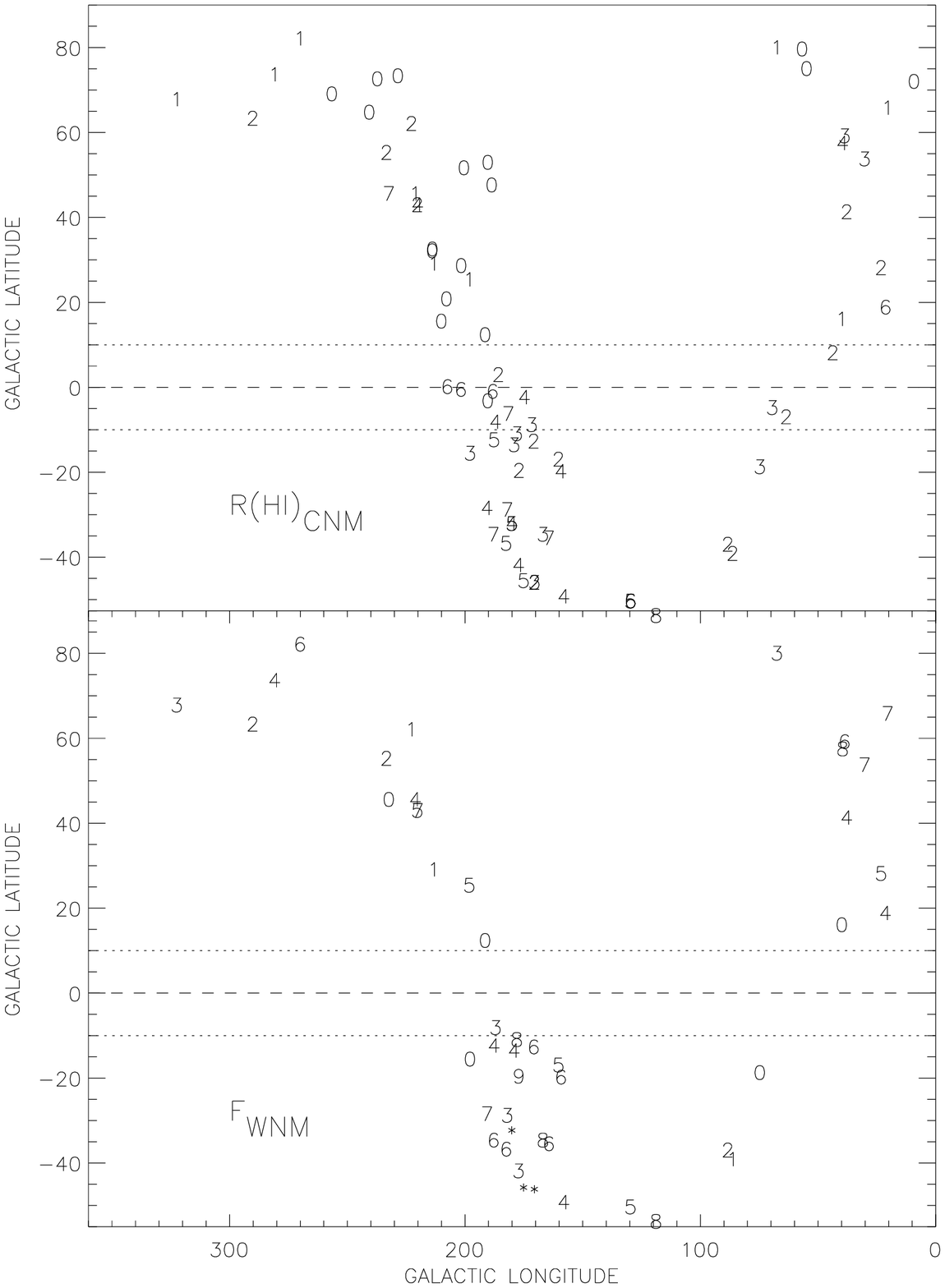} 
\end{center}

\caption{ {\it Top}: Map of $int(10 R(HI)_{CNM})$, the fraction of CNM
to total column density for each line of sight; for example, 0 means
$R(HI)_{CNM}$ lies between 0 and 0.1. {\it Bottom}: For the MO model
fits in \S \ref{momodel}, map of $F_{WNM}$, the column-density fraction
of thermally unstable gas.  \label{mapnhratio} \label{mapmo}} 
\end{figure}

	Figure \ref{nhtotalnhratio} plots $R(HI)_{CNM}$ versus
$N(HI)_{tot,20}$, with diamonds for $|b|>30^\circ$ and plus signs for
$|b|<30^\circ$.  The separate class of points with $\sum N(HI)_{CNM}=0$
is again distinct and mostly has small $N(HI)_{tot}$.  Apart from this,
a fairly apparent trend is the increase of $R(HI)_{CNM}$ with
$N(HI)_{tot}$ up to a limiting $N(HI)_{tot,20} \sim 12$.  Surprisingly,
this trend levels off, and even seems to reverse, at larger
$N(HI)_{tot}$.  The points following this reversed trend all lie in the
Taurus/Perseus region, where large dust/molecular clouds exist (Figure
\ref{mapnhratio}, top). 

%\clearpage

%\clearpage

%\clearpage

	Figure \ref{mapnhratio} (top) shows a map of $R(HI)_{CNM}$ in
Galactic coordinates.  Points with large and small values of
$R(HI)_{CNM}$ tend to cluster.  In particular, all but three of the
$R(HI)_{CNM}=0$ points fall in Galactic quadrants 3 and 4 $(\ell >
180^\circ, b > 10^\circ)$; this entire region has small values except
for the single isolated, unusual point at $(l, b)=(232^\circ,
47^\circ)$.  This source, which is 3C237, has one component with $\tau_0
= 0.005$ (which is very small) and $T_s = 656$ K (which is the highest
in the sample); it just missed being classed as WNM.  If it had been
classed as WNM, then 3C237 would have had $R(HI)_{CNM}=0.30$ and the
anomaly would be much less severe.  The other three $R(HI)_{CNM}=0$
points cluster with two others with $R(HI)_{CNM}=1$ in the upper right
of the map. 

        We conclude that quadrants 3 and 4, and also the upper right of
the map of Figure \ref{mapnhratio} (top), are definitely unusual in
having very low fractions of CNM.  Both of these regions are disturbed
by supershells.  Heiles (1998) considers the HI, IR, nonthermal radio
continuum, and soft X-ray data and concludes that this general region
has been cleared out by a huge superbubble designated GSH 238+00+09,
powerful enough to have induced the first stages of star formation in
the Vela and Orion regions.  Haffner, Reynolds, and Tufte (1998) have
discovered a huge H$\alpha$-emitting filament that lies in this general
region, which may be part of the same superbubble and also related to
the unusual values for $R(HI)_{CNM}$.  The upper right of the map lies
within the North Polar Spur, a supershell produced by multiple
supernovae in the Sco/Oph star association (Egger 1998). 

\section{THE VOLUME FILLING FRACTION OF THE WNM}

\label{volfill}

	The WNM constitutes about $61\%$ of the total HI column density
for $|b|>10^\circ$ ($\langle R(HI)_{WNM} \rangle = 0.61$ (\S
\ref{cnmwnmsummary}). From large-scale sky surveys the total HI column
density, WNM and CNM combined, follows

\begin{equation}
N(HI)_{20} \sim {3.7 \over \sin |b|}
\end{equation}

\noindent (Kulkarni \& Heiles 1987). Blindly applying our $61\%$ WNM
fraction, we obtain for the typical WNM column density 

\begin{equation}
N(HI)_{WNM,20} \sim {2.1 \over \sin |b|}
\end{equation}

\noindent To progress further we need to adopt a typical temperature for
the WNM. From Figure \ref{histoplot4new}, we use 4000 K; this is simply
an eyeball estimate of a reasonable value for the purpose of the
immediate discussion and is not a median or mean. If the WNM is in
pressure equilibrium with the CNM, with $P/k = 2250$ cm$^{-3}$ K
(Jenkins \& Tripp 2001), then its typical volume density is $n(H)_{WNM}
\sim 0.56$ cm$^{-3}$. Similarly, with the typical CNM temperature of 40
K, the typical CNM volume density is $n(HI)_{CNM} \sim 56$ cm$^{-3}$.
With $\langle R(HI)_{WNM} \rangle = 0.61$, the WNM has about 1.5 times
more mass than the CNM and the WNM occupies 150 times more volume than
the CNM. These ratios are based on the total column density at $|b| >
10^\circ$ and covers all $z$ heights.

	We cannot specify a volume filling fraction for the WNM because
our observations are concentrated at $|b|>10^\circ$ where our lines of
sight extend through the top of the gas layer. The total interstellar
pressure drops by $30\%$ to $40\%$ from $z=0$ to 200 pc (Boulares \& Cox
1990), so one expects on theoretical grounds that the WNM fraction
should increase with $z$.

	We can estimate the volume filling fraction for $z=0$.  However,
doing so requires knowing $\langle n(H_2) \rangle $, the mean H$_2$
volume density at $z=0$. This is uncertain because it depends on
converting CO profile areas to H$_2$ column densities, which relies on
the so-called $X$ factor. Dame et al (1987) used $X=2.7 \times 10^{20}$
cm$^{-2}$ K km s$^{-1}$ to obtain  $\langle n(H_2) \rangle = 0.14$
cm$^{-3}$; correcting this for the more recent $X=1.8 \times 10^{20}$
cm$^{-2}$ K km s$^{-1}$ found by Dame, Hartmann, \& Thaddeus (2001)
gives $\langle n(H_2)\rangle = 0.09$ cm$^{-2}$. Solomon (personal
communication) estimates $\langle n(H_2)\rangle \approx 0.47$ cm$^{-3}$
and Blitz (personal communication) estimates $\langle n(H_2)\rangle
\approx 0.25$ cm$^{-3}$. We will use the mean of these three numbers,
which is 0.27 cm$^{-3}$, but this is clearly very uncertain. This
corresponds to a total H-nuclei column density of $16.7 \times 10^{20}$
cm$^{-2}$ kpc$^{-1}$. 

	We can now estimate the volume filling fraction for $z=0$.  At
$z=0$ the reddening is $\sim 0.53$ mag kpc$^{-1}$, which corresponds to
$(N(HI) + 2N(H_2))_{20} = 31$ kpc$^{-1}$ (Binney \& Merrifield 1998). 
Of this, the H$_2$ contributes  $16.7 \times 10^{20}$ per kpc, leaving
$14.3 \times 10^{20}$ cm$^{-2}$ kpc$^{-1}$ for HI.  From \S
\ref{cnmwnmsummary}, we will adopt the tentative $|b|<1.3^\circ$ value
$\langle R(HI)_{WNM} \rangle = 0.61$; thus $N(HI)_{WNM,20} \sim 8.7$
kpc$^{-1}$, which  corresponds to $\langle n(HI)_{WNM}
\rangle = 0.28$ cm$^{-3}$. With a true volume density of 0.56 cm$^{-3}$,
the WNM volume filling fraction $\sim 0.50$. 

	Our WNM filling factor, $\sim 0.50$, includes the HI in
partially ionized Warm Ionized Medium (WIM) and is therefore larger than
the filling factor of the WNM alone. This makes it quite close to the
filling factor derived by MO, whose corresponding value is $\sim 0.40$
at $z=0$.

	This WNM volume filling fraction at $z=0$, $0.50$, is {\it very
rough} because of uncertainties in the following: the accuracy of our
low-latitude data; the typical WNM temperature (which we took as 4000
K); the Jenkins \& Tripp CNM pressure (Wolfire et al 2002), which we
used also for the WNM pressure; the WNM volume density, which is derived
from the aformentioned WNM density and temperature; the reddening per
kpc; the $X$ factor; and the mean CO profile area in the Solar vicinity.
Moreover, it may not apply elsewhere if the Solar vicinity is unusual.
In the nearby Solar vicinity the most of the remaining volume is
probably occupied by the superbubble HIM as cataloged and crudely
sketched by Heiles (1998). The nearby Solar vicinity may have an
unusually large fractional volume filled by superbubbles because the
average over the disk should be about 0.1 (McKee 1993). 

\section{STATISTICS ON $V_{LSR}$}

\label{vlsrstatistics}

%\clearpage
\begin{figure}[h!]
\begin{center}
\includegraphics[width=3.5in] {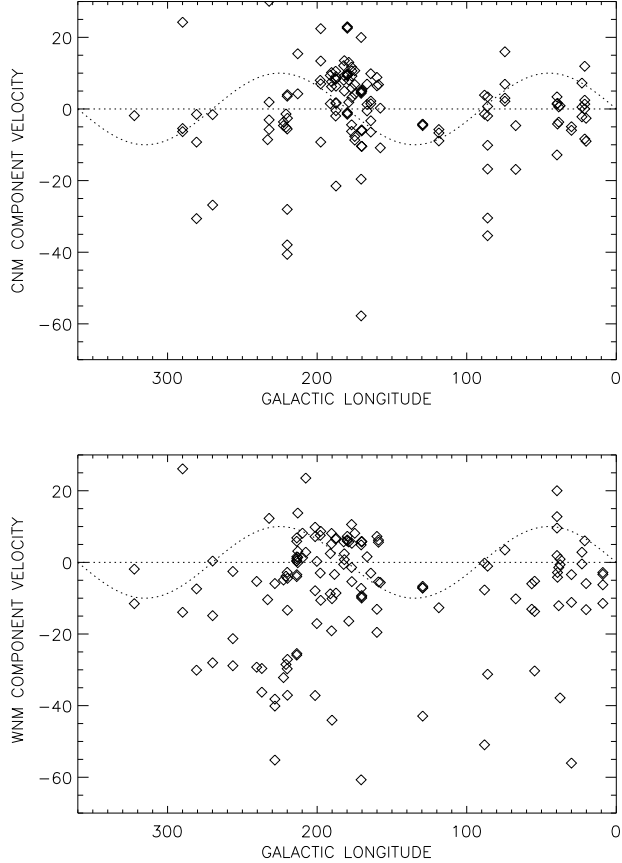}
\end{center}

\caption{ $V_{LSR}$ versus Galactic longitude for CNM components (top)
and WNM components (bottom), for sources with $|b|>10^\circ$. The dotted
line indicates Galactic rotation with an arbitrary amplitude of 10 km
s$^{-1}$. \label{vcenplotnhgaussians}}  \end{figure}
%\clearpage

        With a good sampling of the sky one could use our Gaussian
$V_{LSR}$'s and Galactic rotation to determine the mean scale heights of
the WNM and CNM. However, Arecibo's restricted declination coverage
makes our sky coverage too poor for this purpose. Figure
\ref{vcenplotnhgaussians} shows $V_{LSR}$ versus $l$ for the CNM (top)
and WNM Gaussians, together with a 10 km s$^{-1}$ sinusoid to illustrate
the expected algebraic sign versus $l$ (the expected amplitude is much
smaller). The points exhibit a huge scatter and no tendency to change
sign in the expected way. Galactic rotation contributes no recognizable
signature to the component velocities.

%\clearpage
\begin{figure}[h!]
\begin{center}
\includegraphics[width=3.5in] {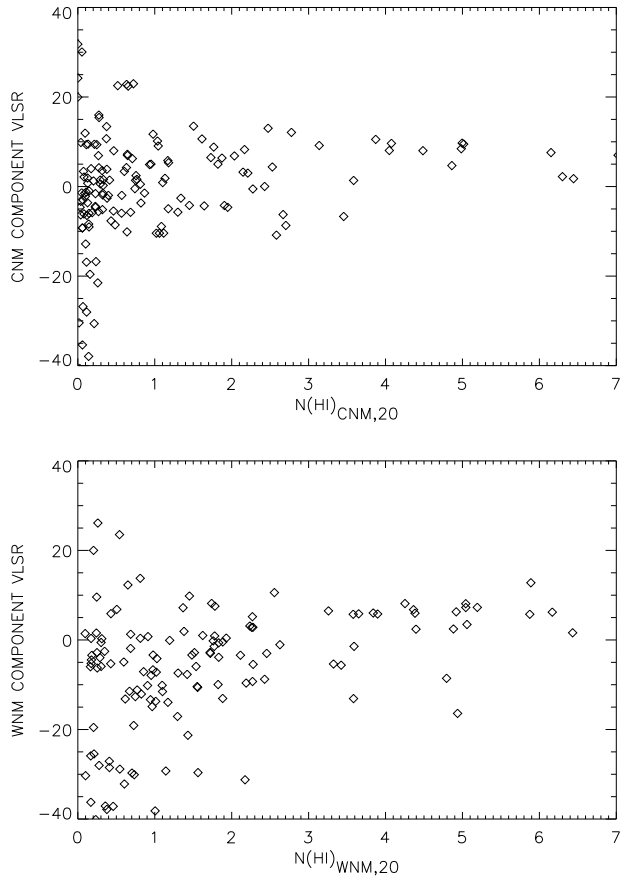}
\end{center}

\caption{ $V_{LSR}$ versus $N(HI)_{20}$ for CNM components (top) and WNM   
components (bottom), for sources with $|b|>10^\circ$. $N(HI$ is in units
of $10^{20}$ cm$^{-2}$.
\label{vcenplotnhgaussians2}} \end{figure}
%\clearpage

	The standard deviations of the Gaussian component center
velocities (i.e., on a component-by-component basis) for the (CNM, WNM)
are $\sigma_{VLSR} = (14.0,16.1)$ km s$^{-1}$.  Weighted by column
density, these become $\sigma_{VLSR} = (7.1,11.4)$ km s$^{-1}$; the
smaller values reflect the fact that higher column density components
have smaller $\sigma_{VLSR}$, as shown in Figure
\ref{vcenplotnhgaussians2}.  These column-density-weighted values
correspond to FWHM $\Delta V_{FWHM} =(16.6, 26.9)$ km s$^{-1}$ and to
$T_{kmax}= (6000, 15900)$ K. The CNM $\sigma_{VLSR}$ is somewhat larger
than the typical WNM sound velocity, indicating that if the CNM consists
of clumps moving within a substrate of WNM, then that motion is mildly
supersonic unless, perhaps, the WNM is permeated by a magnetic field.

\section{ RELATIONSHIPS AMONG LOGARITHMS OF $T_s$, $\tau_0$, $N(HI)$,
AND $T_{kmax}$ FOR THE CNM COMPONENTS}

\label{correlations}

	In this section we discuss correlations among the logarithms of
the four CNM parameters $(T_s, \tau_0, N(HI), T_{kmax})$.  Significant
correlations exist among all pairs of parameters. This is most easily
shown in the correlation matrix:

\begin{eqnarray} \label{corrmatrixeqn}
\left[
\begin{array}{cccc}
1.00    & 0.69 & 0.31 & -0.53 \\
0.69 & 1.00    & 0.38 & -0.40 \\
0.31 & 0.38 & 1.00    & 0.59  \\
-0.53&-0.40 & 0.59 & 1.00     \\
\end{array}
\; \right]
\left[
\begin{array}{c}
\log T_s \\
\log T_{kmax} \\
\log N(HI) \\
\log \tau_0 \\
\end{array}
\; \right]
\end{eqnarray}

\subsection{The Historical $\tau_0$-$T_s$ relationship}

	Most previous studies of HI opacity (see review by Kulkarni \&
Heiles 1987) have searched for and found a statistical
relationship between the spin temperature and peak optical depth of the
form 

\begin{equation} \label{ttaueqn}
\log( T_s) = \log(T_{s0}) +  B \log(1 - e^{-\tau_0})
\end{equation}

\noindent where temperatures are in Kelvins and we write the equation to
explicitly emphasize that the least squares fits are done to the
logarithms of the data, not the data.  Typically these studies find 
$[T_{s0}, B] \sim [60 \ {\rm K}, -0.35]$. The (improper: see below) fit
for our data is not dissimilar, yielding $[T_{s0}, B] = [(33 \pm 4) \
{\rm K}, (-0.29 \pm 0.05)]$ (we fit $\log( T_s)$ to $\log (\tau_0)$
instead of to $\log(1 - e^{-\tau_0})$; the difference is unimportant
because most $\tau_0$ are small). Mebold et al (1982) find no
significant relationship. The form of equation \ref{ttaueqn} has no
physical rationale; it is simply a convenient representation of the
data.  Moreover, $\tau_0$ has no physical influence in the CNM
environment so in no case can we regard equation \ref{ttaueqn} as being
causal. On the other hand, PST and Liszt (1983) discuss physical models,
involving a cold cloud surrounded by a warm envelope, that lead to
reasonable matches with equation \ref{ttaueqn}.

	There are two problems with these historical observational
results for equation \ref{ttaueqn}. One is that the least squares fits
are performed in the conventional way, specifically that the
observational errors in the independent variable $(1 - e^{-\tau_0})$ are
ignored and implicitly set to zero; this always produces too flat an
estimate of the slope (Stetson 2002; Heiles 2002).  Thus the typical
true slope is more negative than $-0.35$. Much more serious is the
presence of the other two parameters $N(HI)$ and $T_{kmax}$.  Our four
parameters exhibit the mutual correlations shown in equation
\ref{corrmatrixeqn}. These mutual correlations render meaningless the 
results of least-squares fits done on only selected pairs of variables.
In particular, equation \ref{corrmatrixeqn} shows that there is no
special significance to the $(\tau_0, T_s)$ pair because other parameter
pairs exhibit similar levels of correlation; the $(\tau_0, T_s)$ pair was
emphasized in earlier studies because they did not use Gaussian
components, so they had no measure of the linewidth $T_{kmax}$ or
$N(HI)$.  

	Even if there were no mutual correlations a $\tau_0$-$T_s$
relationship would occur naturally. Our four parameters are physically
related through the usual equation

\begin{equation} \label{nheqn}
N(HI)_{20} = 0.0195 \tau_0 T_s \Delta V_{FWHM} = 
0.0042 \tau_0 T_s T_{kmax}^{1/2}
\end{equation}

\noindent where $N(HI)_{20}$ is in units of $10^{20}$ cm$^{-2}$ and
$\Delta V_{FWHM}$ is the FWHM in km s$^{-1}$.  If all clouds have the
same or randomly distributed $N(HI)$ and $T_{kmax}$, then we would
expect an inverse correlation between $\tau_0$ and $T_s$ with
logarithmic slope $-1$. When we properly fit this pair of parameters
with our data, accounting for uncertainties in both parameters, we
obtain $[T_{s0},B] = [ (18 \pm 2) {\rm K}, (-0.70 \pm 0.04)]$; the slope
is fairly close to $-1$.\footnote{This slope, $-0.70$, is significantly
steeper than the $-0.29$ derived by ignoring the errors in $\tau_0$, an
illustration of the danger inherent in using inappropriate fitting
techniques.} Clearly, the $\tau_0$-$T_s$ relationship needs to be
considered in the light of a comprehensive multivariate analysis. We
revisit the relationship in this light below in \S \ref{revisit}.

\subsection{ Principal Components Analysis}

	This is a multivariate data set and an appropriate tool for its
investigation is Principal Components Analysis (PCA). For an
$N$-parameter dataset, PCA is a general technique to determine the $N$
different linear combinations of the parameters that express the
characteristics of the data more naturally than do the $N$ parameters
individually. PCA works using the datapoints themselves, without
preconceived notions of what might be significant. Dunteman (1984)
provides a good introduction including a graphical illustration for a
two-parameter example, while Murtagh \& Heck (1987; based on Lebart,
Morineau, \& Warwick 1984) provides a more thorough discussion,
including software. 

\subsubsection{ Quick Description of PCA: the two-parameter example}

	We present a quick description of the idea for the uninitiated
reader. In our case of four correlated parameters, the datapoints fall
in a four-dimensional hyperellipsoid, which is somewhat difficult to
envision, so we describe an example with only two variables $(x,y)$. The
datapoints fall in an ellipse on the $(x,y)$ plane; the principal axes
of the ellipse intersect in a center, and they have an axial ratio and
slope. These axes are eigenvectors that define the two linear and
orthogonal combinations of $(x,y)$ that best represent the datapoint
ellipse. 

	Suppose, as a simple example, that $(x,y)$ represent
(luminosity, color) of stars and we look only at main sequence stars
with zero reddening. Then the datapoints fall on a line, which is the
main sequence, and departures from the line result only from
observational errors, which are small but nonzero. Then the longer
principal axis of the ellipse represents the main sequence, and its
associated eigenvector represents the linear combination of $(x,y)$ that
defines the main sequence. The position along this eigenvector is a
measure of the  stellar mass. The spread (variance) of datapoints along
this line is large and represents the range of stellar masses.  This
illustrates that the eigenvector associated with the largest variance is
the most important. The shorter principal axis represents the
measurement errors, and the variance along this line is small. In this
example, the two eigenvectors have definite and distinct physical
meanings. The specification of these eigenvectors, with their minimum
and maximum variances, is equivalent to a least squares fit; when there
are more than two parameters, PCA automatically extracts the most
significant combinations of parameters (the eigenvectors) for variance
maximization.

	The real difficulty in PCA is the interpretation. One hopes that
the eigenvectors fall into two classes, one with high and one with low
variance. The high-variance classes provide physically significant
combinations of the original parameters. The low-variance classes
provide approximate linear relationships among the original parameters. 

	In particular, an eigenvector with {\it zero} variance reveals
an {\it exact} linear relationship among the parameters. In our case,
the four parameters are rigorously related by equation \ref{nheqn}: the
three parameters on the right hand side are determined observationally,
and $N(HI)$ is derived from them.  Thus the PCA analysis should produce
one eigenvector with zero variance and its linear combination of
parameters should correspond to the logarithmic form of equation
\ref{nheqn}.  Moreover, if we perform a simultaneous least-squares fit
of any one of these four parameters to the other three, we necessarily
recover the dependencies in equation \ref{nheqn}.

	Below we will find that two of our eigenvectors have small
variance. This provides two relationships among the parameters. Of
course, we will also have two eigenvectors with large variance, meaning
that only two linear combinations of parameters are both sufficient and
necessary to specify the physical description of a CNM cloud. Because
the parameters are all related we have our choice regarding how we
actually  express these eigenvectors. 

	If we were to be so fortunate as to find three eigenvectors with
low variance, then three of the four parameters would be expressible in
terms of the fourth, and CNM clouds would be characterized by only a
single eigenvector---a single combination of parameters. In our example
of stars above, this is not the case because other parameters such as
reddening, metallicity, and age also determine the observable properties
of a star. It isn't the case for CNM clouds, either.

	In general, the number of eigenvectors must equal the number of
parameters. PCA extracts the eigenvectors and their associated variances
from the datapoints themselves. For multivariate datasets it is
exceedingly useful for exploring fundamental relationships among the
parameters. However, it is not a panacea. It cannot deal with differing
uncertainties among the datapoints, it cannot derive nonlinear
combinations of the parameters, and it cannot provide uncertainties in
the derived eigenvectors. Below, we use PCA in combination with
least-squares fitting to explore the relationships among our four
parameters.

\subsubsection{ PCA with our four parameters}

	We applied PCA to our datapoints. As is required for physically
meaningful results, we first standardized the measured datapoints by
removing means and forcing variances to be equal. Then we performed the
PCA. Finally, we reversed the standardization procedure so that we could
express the eigenvectors in terms of the original measured parameters.

	Fortunately, the eigenvectors do in fact divide into the two
classes. The two eigenvectors with large variances are

\begin{mathletters} \label{eveqns}
\begin{equation} \label{evone}
EV1: \ \log T_s + 0.74 \log T_{kmax} + 0.09 \log N(HI)_{20} 
	- 0.41 \log \tau_0 - 3.88\ ; \ variance=0.52
\end{equation}
\begin{equation} \label{evtwo}
EV2: \ \log T_s + 1.57 \log T_{kmax} + 4.31 \log N(HI)_{20} 
	+ 2.88 \log \tau_0 - 1.88\ ; \ variance=0.40
\end{equation}
\end{mathletters}

\noindent Here we express variances in fractions of the total, so the
sum of the four adds to unity; also, the lengths of eigenvectors are
arbitrary, and we have arbitrarily made the coefficient of $\log T_s$
equal to unity. For the two eigenvectors having small variances, we set
the eigenvectors equal to zero to provide the corresponding equations
that relate the parameters. This is strictly valid for the eigenvector
$EV4$ with zero variance, but only approximately so for $EV3$:

\begin{mathletters} \label{eveqnssmall}
\begin{equation} \label{evthree}
EV3: \ \log T_s = 0.85 \log T_{kmax} - 0.10 \log N(HI)_{20} 
	+ 0.006 \log \tau_0 - 0.29 \ ; \ variance=0.08
\end{equation}
\begin{equation} \label{evfour}
EV4: \ \log T_s = -0.50 \log T_{kmax} + 1.00 \log N(HI)_{20} 
	- 1.00 \log \tau_0 + 2.38 \ ; \ variance=0.00
\end{equation}
\end{mathletters}

\noindent Equation \ref{evfour} corresponds exactly to equation
\ref{nheqn}. 

	In equation \ref{evthree} we can ignore the tiny coefficient of
$\log \tau_0$, so this equation provides $T_s$ in terms of $[T_{kmax},
N(HI)_{20}]$. This is similar to a least squares fit for $T_s$ in terms
of $T_{kmax}$ and $N(HI)$ (see \S \ref{leastsquaresfits}).
Alternatively, we can extend the $\tau_0-T_s$ relationship to include a
term in $\log (N(HI))$ by using equations \ref{eveqnssmall} to eliminate
$T_{kmax}$:

\begin{equation} \label{evfive}
 \log T_s = 0.59 \log N(HI)_{20} 
	- 0.62 \log \tau_0 + 1.39\ 
\end{equation}

\noindent We hasten to emphasize that we regard this as a mathematical
relationship only with no direct physical significance.

\subsubsection{ The two fundamental CNM eigenvectors: expressible in two
measured parameters} \label{sptpt}

	Finally, we can use equations \ref{eveqnssmall} to eliminate two
parameters from the physically significant eigenvectors in equations
\ref{eveqns} so as to determine combinations of physically significant
cloud parameters. Clearly, $\tau_0$ should be one parameter that is
eliminated because it should have no causal influence. Of the three
remaining ones, we believe that $N(HI)$ should {\it not} be eliminated
because it is a naturally fundamental quantity that determines the
extent to which the cloud interior is shielded from starlight and cosmic
rays.  This leaves us with the choice of eliminating either $T_s$ or
$T_{kmax}$. It isn't clear {\it a priori} which is more physically
important, so we provide two versions of the two eigenvectors. First, in
terms of $(N(HI), T_{kmax})$:

\begin{mathletters} \label{evsix}
\begin{equation}
EV1: \ 0.41 \log N(HI)_{20} -0.91 \log T_{kmax} + 4.43
\end{equation}
\begin{equation}
EV2: \ 0.76 \log N(HI)_{20} + 0.65 \log T_{kmax} + 0.40
\end{equation}
\end{mathletters}

\noindent and next, in terms of $(N(HI), T_s)$:

\begin{mathletters} \label{evseven}
\begin{equation}
EV1: \ -0.08 \log N(HI)_{20} +1.00 \log T_s - 1.79
\end{equation}
\begin{equation}
EV2: \ 0.97 \log N(HI)_{20} - 0.23 \log T_s + 0.68
\end{equation}
\end{mathletters}

\noindent Here we have arbitrarily forced the squares of the
coefficients of $\log N(HI)_{20}$ and $\log T_s$ to sum to unity. 

	Can we interpret these eigenvectors in physical terms? For the
first set in equations \ref{evsix}, $[EV1,EV2]$ correspond approximately
to $[\left( {N(HI) \over \Delta V^4 } \right), N(HI)\Delta V]$. We
discern no physical meaning for $EV1$. In contrast, $EV2$ represents the
total opacity of the cloud to spectral lines, and we have in mind in
particular the CII 157$\mu$m cooling line. 

	For the second set in equations \ref{evseven}, there is a very
straightforward physical interpretation for the eigenvectors. The
differences between the coefficients of $\log N(HI)_{20}$ and $\log T_s$
are large. Roughly speaking, $EV1$ corresponds to $(\log T_s)$ and $EV2$
to $(\log N(HI)_{20})$. In other words, the two eigenvectors can be
taken to be these two parameters instead of two combinations of all four
parameters. Writing the two eigenvectors as $[EV1, EV2]=[ \log (T_s),
\log (N(HI))]$ makes physical sense: $T_s$ makes sense because the CNM
cooling time is short, $\sim 5000$ yr, so the kinetic temperature is a
sensitive indicator of the current balance between heating and cooling
processes; $N(HI)$ makes sense because column density shields the cloud
from the external environment and seems equivalent to mass for a star.
We conclude that these two parameters---kinetic temperature and HI
column density---are convenient, physically meaningful, and
approximately orthogonal ones for CNM components. 

\subsubsection{ Least Squares Fits} \label{leastsquaresfits}

	The relationship of equation \ref{evthree} comes from PCA, not a
least squares fit, so it does not weight datapoints according to their
intrinsic uncertainties. Here we perform least squares fits that remove
this deficiency. We cannot use conventional least squares fitting
because it assumes that the uncertainties in the independent variables
are zero. Accordingly, we generalize Stetson's (2002) technique to
include multiple independent variables; this is discussed in detail by
Heiles (2002). 

	We take the set of three variables $[\log (T_s), \log
(T_{kmax}), \log(N(HI))]$ and perform two independent fits by permuting
the independent and dependent variables. These two different fits
provide identical chi-square and, also,  self-consistent values for the
coefficients and their errors, as they should if the errors in all
parameters are properly treated. The result is

\begin{equation} \label{lsfiteqnoneb}
\log T_{kmax} = (1.14 \pm 0.05) \log T_s + (0.27 \pm 0.05) \log N(HI)_{20} 
	+ (0.31 \pm 0.12)  \ \ ; \ \widehat{\chi}^2=101
\end{equation}

\noindent The reduced chi square $\widehat{\chi}^2 = 101$, which means
individual points depart from the fit by typically 10 times their
intrinsic uncertainties---the fit should be regarded as a trend instead
of an accurate representation of individual datapoints. 

	We go further by exploring the relationship in the form of
equation \ref{lsfiteqnoneb}. We begin our exploration by performing the
fits of $\log (T_{kmax})$ to $\log (T_s)$ and to $\log (N(HI))$
independently. These fits yield

\begin{mathletters} \label{lsfiteqntwo}
\begin{equation} \label{lsfiteqntwoa}
\log T_{kmax} = (1.32 \pm 0.05) \log T_s - (0.11 \pm 0.13)  
	\ \ ; \ \widehat{\chi}^2=117
\end{equation}
\begin{equation} \label{lsfiteqntwob}
\log T_{kmax} = (1.11 \pm 0.08) \log N(HI)_{20} + (3.09 \pm 0.05)  
	\ \ ; \ \widehat{\chi}^2=902
\end{equation}
\end{mathletters}

\noindent The widely different values for $\widehat{\chi}^2$ show that
the latter fit, equation \ref{lsfiteqntwob}, represents the data far
less well than the former. Moreover, $\widehat{\chi}^2$ for equation
\ref{lsfiteqntwoa} is only marginally worse than that for equation
\ref{lsfiteqnoneb}. We conclude that the trend of variation of
$T_{kmax}$ is as well enough expressed by \ref{lsfiteqntwoa}. 

%\clearpage
\begin{figure}[p!]
\begin{center}
\includegraphics[width=5.0in]{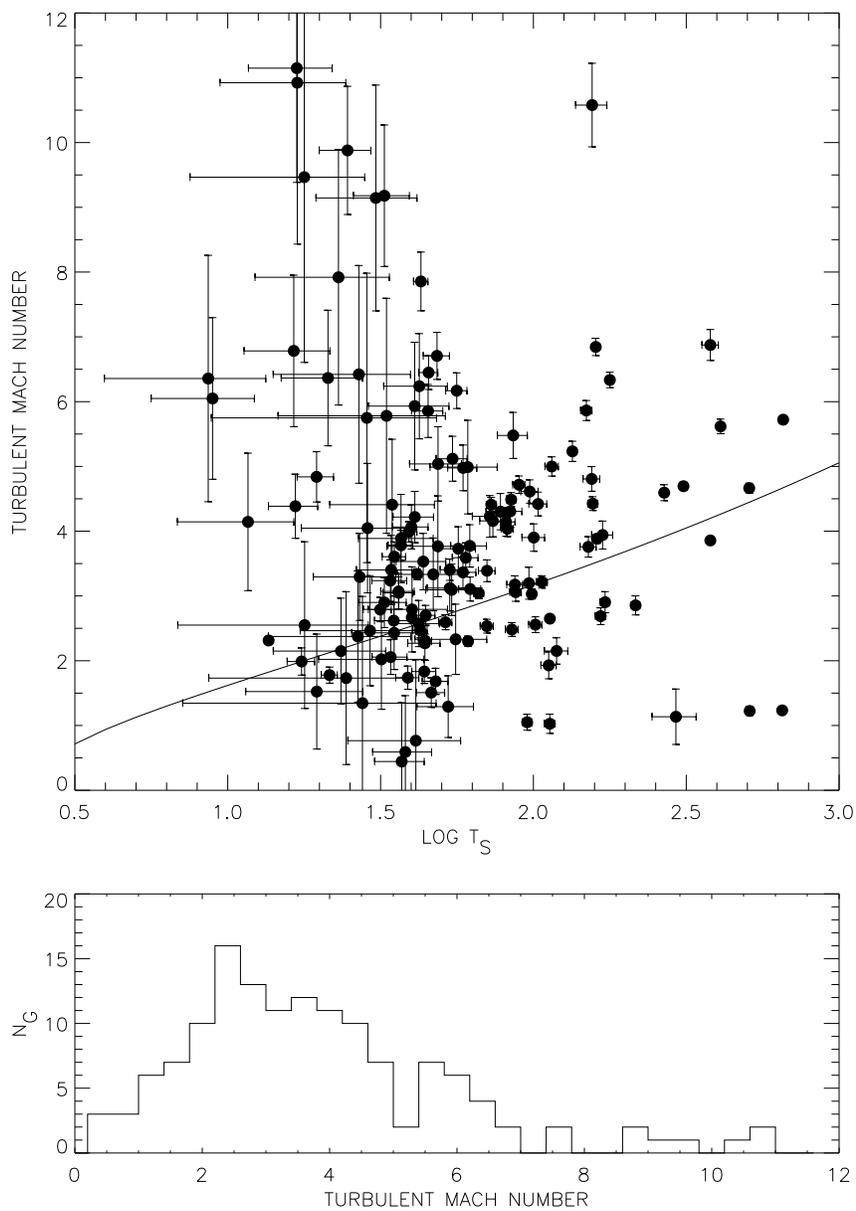}
\end{center}

\caption{ {\it Top:} Turbulent Mach number $M_t$, defined in the text
just above equation \ref{macheqn}, versus $\log T_s$. Errorbars are
$1\sigma$. The solid curve is equation \ref{macheqn} {\it Bottom:}
Histogram of $M_t$ for Gaussian components. \label{lsfitfig}}
\end{figure} 

%\clearpage

        Equation \ref{lsfiteqntwoa} can be written as the ratio
${T_{kmax} \over T_s}= 0.78 T_x^{0.32}$. Also, the ratio ${T_{kmax}   
\over T_s}$ can be used to determine the mean square turbulent  velocity

\begin{equation}
V_{t, 1d}^2 =  {k T_s \over m_H} \left( {T_{kmax} \over T_{s}} -1 \right)
\end{equation}

\noindent Multiplying this by 3 gives the mean square 3-dimensional
turbulent velocity $V_{t, 3d}^2$, and dividing the latter by the square
of the sound velocity $C_s$ gives the square of the turbulent Mach
number $M_t$. The appropriate sound velocity is the isothermal one   
because thermal equilibrium is reached quickly in the CNM. We adopt a
mean atomic weight of $1.4 m_H$, corresponding to a fractional He
abundance of 0.15 by number so that $C_s^2= {k T_s \over 1.4 m_H}$. With
this,

\begin{equation}
M_t^2 = {V_{t, 3d}^2 \over C_s^2} =
  4.2  \left( {T_{kmax} \over T_{s}} -1 \right)
\end{equation}

\noindent Using equation 15a for the fit to the typical temperature
ratio, we have

\begin{equation} \label{macheqn}
M_t \sim 3.3 (T_{s,40}^{0.32} - 0.40)^{1/2}
\end{equation}

        The top panel of figure \ref{lsfitfig} shows the datapoints   
together with this fit. There is much scatter, which is exacerbated by
the errors on the measured quantities. Despite the perhaps disappointing
visual appearance, most of the points do fall fairly close to the line,
as revealed by the histogram of $M_t$ in the bottom panel. Thus, very
roughly speaking, the internal CNM macroscopic nonthermal motions are
characterized by Mach number $\sim 3$; CNM clouds tend to be strongly   
supersonic. Individual components differ markedly from this value and
there is a weak systematic increase with $T_s$.
        
\subsection{The $\tau_0$-$T_s$ relationship revisited---and
relinquished} \label{revisit}

	Here we revisit the $\tau_0$-$T_s$ relationship by discussing
least squares fits on the various parameter combinations ($T_s, N(HI),
\tau_0)$. The results of these fits are:

\begin{mathletters}
\begin{equation}
\log T_s = (1.25 \pm 0.06) - (0.70 \pm 0.04) \log \tau_0 \ ; \ 
	\widehat{\chi}^2 = 141
\end{equation}
\begin{equation} \label{seventeenb}
\log T_s = (2.45 \pm 0.03) + (0.60 \pm 0.07) \log N(HI)_{20} ; \ 
	\widehat{\chi}^2 = 532
\end{equation}

\noindent (The high $\widehat{\chi}^2$ for equation \ref{seventeenb} is
another indication that $\log(T_s)$ and $\log(N(HI))$ are approximately
orthogonal, just as we conclude from the PCA in section \ref{sptpt}.) 

\begin{equation} \label{seventeenc}
\log T_s = (1.39 \pm 0.01) - (0.64 \pm 0.01) \log \tau_0 + 
	(0.57 \pm 0.01) \log N(HI)_{20} ; \ \widehat{\chi}^2 = 7.5
\end{equation}
\end{mathletters}

\noindent In contrast to the situation of \S \ref{leastsquaresfits}, the
fit that includes both $\tau_0$ and $N(HI)$ provides a far smaller
$\widehat{\chi}^2$ than does either of the single-parameter fits.
Moreover, the parameters are very well determined. This means that $T_s$
is {\it not} well-predicted by only $\tau_0$, as is expressed in the
classical $\tau_0$-$T_s$ relationship. 

	This good fit of equation \ref{seventeenc} is primarily a matter
of two relationships: \begin{enumerate}

	\item Equation \ref{lsfiteqntwoa}, which is approximate. It
relates $\log (T_s)$ and $\log (T_{kmax})$ in an approximately linear
fashion, meaning that these two parameters are highly correlated so that,
in a least squares fit that included both, the pair would be nearly
degenerate. Thus, eliminating $\log (T_{kmax})$ as an independent
variable in a least squares fit has little effect on the quality of a
fit for $\log (T_s)$. 

	\item The logarithmic form of equation \ref{nheqn}, which is
exact and would produce a perfect least squares fit for $\log (T_s)$ if
we included the three other parameters. Not including $\log (T_{kmax})$,
which is nearly degenerate with $\log (T_s)$, makes the fit only very
good instead of perfect.

\end{enumerate}

	We conclude that there is no physically significant
$\tau_0$-$N(HI)$-$T_s$ relationship, except as related through equation
\ref{nheqn}. 

\section{RAMIFICATIONS OF ISOTROPIC CNM CLOUDS AT KNOWN PRESSURE}

\label{pequality}

	Here we discuss the effect of inadequate angular resolution
(``beam dilution'') on our derived CNM spin temperatures and column
densities.  We derive physical sizes of CNM components by assuming that
the pressure is known. CNM pressures have been measured by Jenkins \&
Tripp (2001), who find a histogram that peaks near $(P/k) = nT = 2250$
cm$^{-3}$ K, with wide tails. Here we will normalize the ISM pressure in
these units, i.e. we write $(P/k) = 2250 P_{2250}$, and normalize the
measured temperatures in units of $T_{s,40}=40$ K, which is close to our
histogram peak (Figure \ref{histoplot4new}.  We will denote true
quantities with a superscripted $^*$ and the observed ones with no
superscript. For example, the observed spin temperature is $T_s$ and the
true one is $T^*_s$.

\subsection{Volume density and size under pressure equality} \label{isotropic}

	The column density $N(HI)$ of a Gaussian provides no information
on its volume density $n(HI)$ or linear size $L$. We can obtain these
quantities if we know the pressure. Using the parameterization described
immediately above, we obtain for the volume density

\begin{mathletters}

\begin{equation}
n(HI) = 56 { P_{2250} \over T^*_{s,40} } \ {\rm cm}^{-3} ; 
\end{equation}

\noindent for the length of the cloud along the line of sight

\begin{equation}
L_{||} = 0.57 {T^*_{s,40} \over P_{2250}} N^*(HI)_{20} \ {\rm pc};
\end{equation}

\noindent and, assuming an isotropic cloud, for the angular size, 

\begin{equation} \label{thetaeqn}
\theta_\perp = 20 {T^*_{s,40} \over D_{100} P_{2250}} N^*(HI)_{20} 
   \ {\rm arcmin}.
\end{equation}

\noindent where we normalize the distance to units of 100 pc because
this is the approximate scale height of the CNM (Kulkarni \& Heiles
1987). Actual distances vary widely; for example, the Taurus complex 
has distance 140 pc (Arce \& Goodman 1999) while the nearby Perseus
complex has distance 334 pc (Ladd, Myers, \& Goodman 1994). 

\end{mathletters}

\subsection{ Beam dilution and our derived Gaussian parameters}
\label{drastic}

	If CNM clouds are isotropic, then those with smallest $N(HI)$
will also have the smallest angular sizes as in equation \ref{thetaeqn}.
These same small clouds may also suffer from beam dilution. Therefore,
they will contribute less antenna temperature to our expected emission
profiles, and we will derive values of $T_s$ that are too small. This
effect can lead to a spurious positive correlation between $T_s$ and
$N(HI)$.

	Figure \ref{histoplotnhgaussians} shows that $N(HI)_{CNM,20}$
usually lies in the approximate range 0.03 to 1, corresponding to
$\theta_\perp \sim$ 0.3 to 17 arcmin.  The smaller values violate the
assumptions inherent in our WNM Gaussian fitting process of Paper I \S
4.3, where we assume that the CNM clouds contribute to the expected
profile with no beam dilution---i.e., we assume that they are large
enough to fill the telescope beam of angular diameter $\sim 3.3$ arcmin
(and, more stringently, to fill the beam in the off-source positions,
which lie up to 4.5 arcmin away).  

	To understand this influence, let $N^*(HI)_{CNM}$ and $T^*_s$ be
the true values, which are larger than our derived values because of
beam dilution. There are two contributions to beam dilution:
\begin{enumerate}

	\item The ordinary beam dilution that occurs when observing a
source whose diameter is smaller than the beam diameter.  We express
this by the factor $F_B$, i.e.\ the factor by which the antenna
temperature is reduced by the beam dilution.  It obeys (e.g.\ Rohlfs \&
Wilson 2000)

\begin{equation} \label{ten}
F_B \sim  {(\theta_\perp/\theta_H)^2 \over 1 + (\theta_\perp/\theta_H)^2} 
\end{equation}

\noindent where $\theta_H$ is the {\it effective} HPBW.

	\item The increase in effective HPBW caused by our use of
off-source observations to define the cloud's antenna temperature.  This
is fully discussed in Paper I, \S 3.5.  If a cloud is larger than
Arecibo's 3.3 arcmin beam but smaller than the angular offsets for the
off-source spectra, then the derived emission antenna temperature from
the cloud is too small.  This use of off-source data increases the
innate 3.3 arcmin HPBW to the effective one.  This effective HPBW
should be roughly equal to the square root of the sum of the squares of 
the innate HPBW and the angular displacement of the off-source positions
(about 5 arcmin). That is, the effective HPBW is about 5.7 arcmin.
Accordingly, we define the effective HPBW to be

\begin{equation} \label{eleven}
\theta_H = 5.7 F_H \ {\rm arcmin} \ .
\end{equation}

\noindent where $F_H$ is a factor, close to unity, that more exactly
defines the correct effective HPBW. $F_H$ depends on things such as the
exact cloud shape and the intensity distribution within the cloud
boundary. 

\end{enumerate}

	By combining equations \ref{thetaeqn} and \ref{eleven}, we find
that beam dilution becomes significant for $(\theta_\perp / \theta_H)
\lesssim 1$, which occurs for $N^*(HI)_{CNM} \lesssim 0.3 {D_{100}
P_{2250} F_H \over T^*_{s,40}}$.  For this case, we simplify the
following equations by substituting for equation \ref{ten} the much
simpler equation

\begin{equation}
F_B \sim (\theta_\perp/\theta_H)^2 \ .
\end{equation}

\noindent In terms of physical quantities of equation \ref{thetaeqn}
this becomes

\begin{equation} \label{beamdilution} 
F_B \sim \left( {T^*_{s,40} \over
D_{100} P_{2250} F_H} {N^*(HI)_{CNM,20} \over 0.3}
   \right)^2
\end{equation}

\noindent The observed spin temperature $T_s$ and column density
$N(HI)_{CNM,20}$ are both directly proportional
to the antenna temperature, so they are reduced by the same factor:

\begin{mathletters} \label{tsnhfb}
\begin{equation} \label{tsfb}
T_s = F_B T^*_s
\end{equation}
 
\begin{equation} \label{nhfb}
N(HI)_{CNM,20} = F_B N^*(HI)_{CNM,20}
\end{equation}
\end{mathletters} 

\noindent Combining the previous three equations, we obtain $F_B$ in
terms of observed instead of true parameters

\begin{equation} \label{beamdilutioninv} 
F_B \sim \left( {T_{s,40} \over
D_{100} P_{2250} F_H} {N(HI)_{CNM,20} \over 0.3}
   \right)^{2/5}
\end{equation}

	Suppose, for purposes of illustration, that all correlation
coefficients are zero except between $T_s$ and $N(HI)$. Then a least
squares fit between these is meaningful and produces the result $\log
T_s = 2.45 + 0.60 N(HI)_{20}$, i.e.\ $T_{s,40} = 7.0
N(HI)_{20}^{0.60}$. This is, of course, a relation between the {\it
observed} parameters. Using this observed relation together with
equations \ref{beamdilution} and \ref{tsnhfb} to express a new relation
in terms of the {\it true} parameters, we obtain

\begin{equation} T_{s,40}^* = 1.7 (D_{100} P_{2250} F_H)^{0.44}
     N^*(HI)_{CNM,20}^{-0.11} 
\end{equation}

\noindent It is surprising to see that, while the {\it observed} relation
has a {\it positive} slope, the {\it true} relation has a {\it negative}
slope. This illustrates that beam dilution is important and can
drastically affect the relationships among observed quantities.
Historical studies that obtained expected profiles with larger telescope
beams than Arecibo's include Lazereff (1975) and Mebold (1982).

	However, we emphasize that beam dilution effects are much less
severe than we calculate here.  Recall that our analysis applies only in
the case $N^*(HI)_{CNM,20} \lesssim 0.3 {D_{100} P_{2250} F_H \over
T^*_{s,40}}$, a criterion based on the assumption of isotropic clouds
expressed quantitatively in equation \ref{thetaeqn}.  However, we argue
in \S \ref{againstraisin} that CNM components are sheetlike, not
isotropic.  Therefore, they are much more extended in the plane of the
sky than predicted by equation \ref{thetaeqn}, and beam dilution effects
are correspondingly much smaller. 

\section{ EVIDENCE AGAINST ISOTROPIC CNM CLOUDS}

\label{againstraisin}

	If CNM clouds are isotropic, then we predict in \S
\ref{isotropic} (equation \ref{thetaeqn}) the approximate angular size
of CNM clouds.  In particular, all values for $T_s$ and $N(HI)_{CNM}$
are affected by beam dilution when $N^*(HI)_{CNM,20} \left({T^*_{s,40}
\over D_{100} P_{2250}}\right) \lesssim 0.3$.  Here we test this
prediction using five pairs of our sources that are closely-spaced and,
also, using previous observations in the literature.  We find that CNM
clouds are extended over much larger angles than predicted by equation
\ref{thetaeqn}.  Indeed they are often so extended that they appear much
more sheetlike than isotropic. 

\subsection{Evidence from our own data}

	Table \ref{chtbl} lists CNM Gaussian parameters of common
components for our five closely-spaced source pairs.  For each pair, the
parameters for each source are given in fractional form together with
the ratio.  For each individual Gaussian component, we list the derived
$N(HI)_{CNM}$ and also the area under the Gaussian function fit to the
opacity profile.  We believe it is better to compare profile areas.  The
area is derived directly from the opacity profile, while $N(HI)_{CNM}$
is less accurate because it contains the error in derived $T_s$, which
contains the error obtained from combining the opacity profile and the
expected profile.  The expected profile is subject to the additional
uncertainties discussed in Paper I \S 5.2.  These are particularly
serious for weaker opacity components, which are just the ones we are
interested in. 

	For the pair 3C225a/3C225b we list only the strongest opacity
component. From the opacity profiles, one sees that both sources have
two much weaker components in common centered near $V_{LSR}=(-5.6,-2.8)$
km s$^{-1}$. These were included in our fit for 3C225b but not for
3C225a because of the large uncertainty in the opacity profile for
3C225a and our criteria for fitting Gaussians explained in Paper I
\S 5. Visually the two opacity profiles look similar, and if we
had included them it would bolster our case that opacity components
don't change rapidly with position. 

	Scanning Table \ref{chtbl}, we see no little tendency for ratios
to depart from unity more with decreased column density.  Moreover, even
for low column densities the components not only exist for both pair
members, but the ratios usually don't depart too far from unity---only
two of the AREA ratios exceed 2.  Even for the 3C310/3C315 pair,
separated by nearly $2^\circ$, the ratios are quite close to unity. 
This is contrary to the basic prediction of the raisin pudding model. 
The 3C225a/b pair, and in addition the source 3C237, constitute a
special case. 

\subsection{ The 3C225a,b and 3C237 ``Triad Region''}

\label{triadregion}

%\clearpage
\begin{figure}[p!]
\begin{center}
\includegraphics[width=5in] {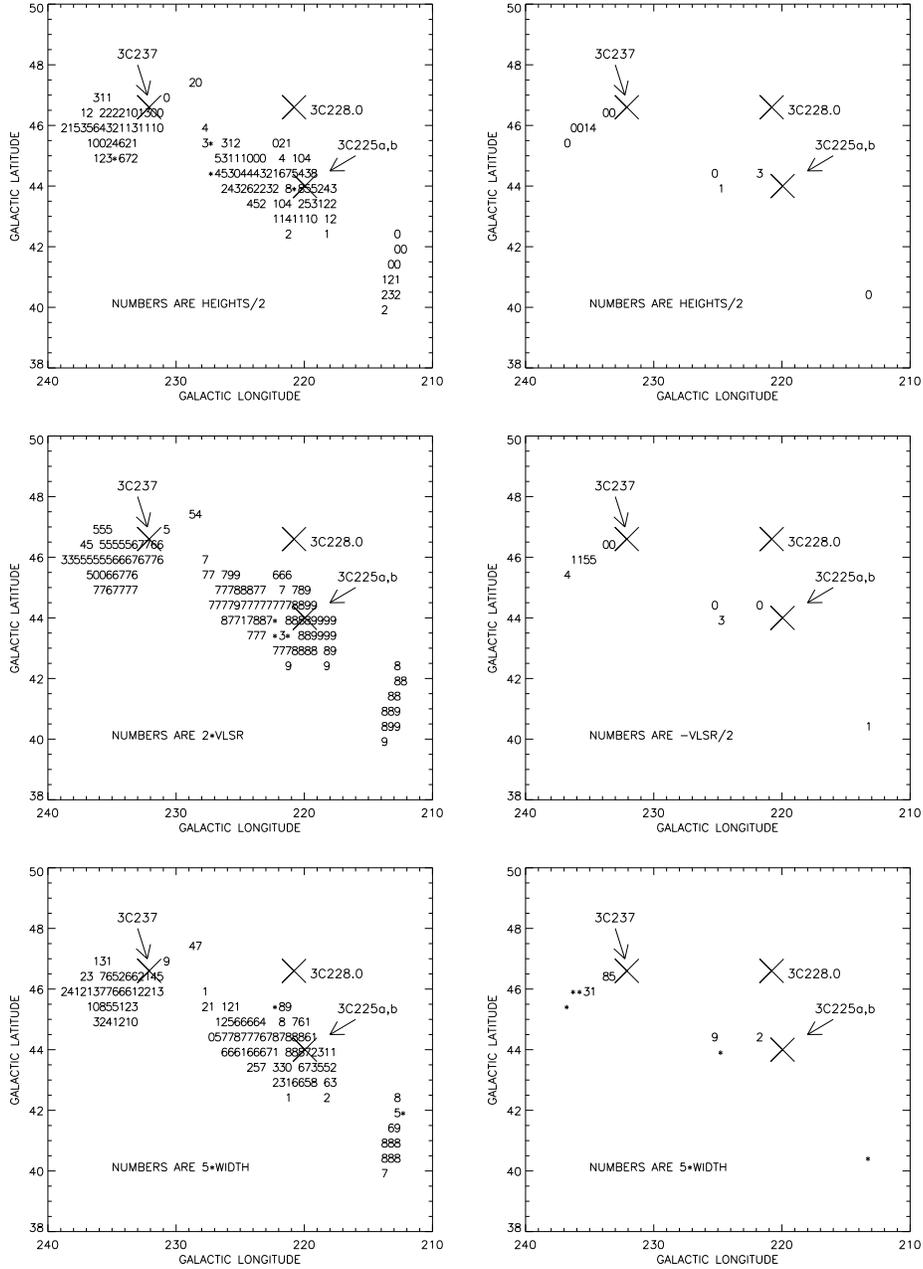} 
\end{center}
\caption{Maps of cold gas ($T_s \sim 25$ K) towards $(l,b) = (225^\circ,
44^\circ)$. Left column shows the predominant component at $V_{LSR} \sim 4$
km s$^{-1}$ and the right the less predominant one at $V_{LSR} \sim -4$ km
s$^{-1}$. The top row shows Gaussian peak height, the center row the
$V_{LSR}$, and the bottom row the halfwidths. Each single-digit number
represents a half-degree pixel. \label{tstmap1}} 
\end{figure}
%\clearpage

	The source pair (3C225a, 3C225b) and also 3C237 all have a
similar opacity component which is prominent in the expected and the
surrounding emission profiles, offering us a unique opportunity to map a
CNM opacity component.  3C237 is about $9^\circ$ away from the 3C225a,b
pair.  The 3C225a,b component was noticed long ago and partially mapped
in emission as ``Cloud A'' by Knapp and Verschuur (1972).  The region
they mapped shows an elongated cloud, at least $5^\circ$ long and about
$1^\circ$ wide; they didn't complete the map and, in particular, didn't
carry it far enough north to include 3C225a and 3C225b.  They derived
the spin temperature by assuming the intrinsic line shape to be Gaussian
and fitting for the saturation; they found $T_s \approx 24$ K over the
whole of the cloud.  They also mapped ``Cloud B'', which is associated
with 3C237 and is similarly cold. 

	In fact, components similar to these clouds exist over tens of
square degrees in the region centered near $(l,b)=(225^\circ,44^\circ)$.
We used the Leiden-Dwingeloo survey of Hartmann and Burton (1997) to map
this feature, exploring the entire positive-latitude region within the
range $l=200^\circ$ to $240^\circ$.  On line profiles, the feature is a
sharp narrow peak on the side of a much broader one, as in our expected
profiles for these sources. To locate positions containing the feature,
we sharpened each profile by subtracting from each profile its 3-point
median-filtered counterpart. Then we least-squares fit the narrow
feature plus a constant and slope and rejected solutions having small
slopes and large widths. We confirmed the suitability of this procedure
by visual inspection of the profiles. 

	Figure \ref{tstmap1} exhibits the results.  Within the region
surveyed we found the feature to exist only within the smaller region
shown; it is possible that the feature extends beyond $l=240^\circ$. The
left column of panels are maps of the three Gaussian parameters height,
center velocity, and width for this cloud, which has $V_{LSR}$ decreasing
slowly from $\sim 4$ to 2 km s$^{-1}$ as $l$ increases to the left
across the map. The right column shows a much less predominate but
similar feature which has $V_{LSR}$ decreasing from $\sim 0$ to $-8$ km
s$^{-1}$ as $l$ increases from the map center towards the left. 

	The predominant component appears as three clouds forming a
ribbon of width $\sim 2^\circ$ and length $\gtrsim 20^\circ$. While the
ribbon is interrupted by voids, the coherence of its characteristics
indicates strongly that it is the really the same physical feature. With
its temperature $T_s \sim 25$ K and typical $N(HI)_{CNM,20} \sim 0.3$, the
volume density $n(HI) \sim 90 P_{2250}$ cm$^{-3}$ and $L_{||} \sim
(0.11/ P_{2250})$ pc. In contrast, the length across the line of sight
is $L_\perp \sim 30 D_{100}$ pc. The aspect ratio is

\begin{equation}
{L_\perp \over L_{||}} \sim 280 D_{100} P_{2250}
\end{equation}

\noindent which is comparable to the aspect ratio for, say, an
old-fashioned LP record. If the Wolfire et al (2002) estimate ${P \over
k} = 3000$ cm$^{-3}$ K is correct, then the ratio is even higher.

	The occasional presence of the negative-velocity feature is
intriguing.  The velocity difference at the positions where it exists
$\sim 5$ km s$^{-1}$. If these two features were the opposite
sides of an expanding shell, then the expansion velocity would be too
small to create a shock in WNM gas. Moreover, both $V_{LSR}$'s are small.
If the feature had been produced by a higher-velocity shock
and slowed to its current $V_{LSR}$, then inhomogeneities in the ISM would
produce significant variations in the current $V_{LSR}$, which don't exist. 
It seems unlikely that the sheetlike structure results from a
shock front. 

\subsection{The ``Small Region'' of Heiles (1967)}

	For his particularly well-conceived thesis, Heiles (1967) used
the mighty NRAO 300-foot telescope to map the 21-cm line in a $\sim 160$
deg$^2$ region centered on $(l,b) \sim (120^\circ, 15^\circ)$. The HI
profiles in this region are characterized by two narrow peaks sitting on
a broad underlying component. In some of the region, the underlying
component has $T_{kmax} \sim 2500$ K. For the two peaks, he estimated
the $\Delta V_{FWHM}$ to be $\sim 3.3$ km s$^{-1}$, which corresponds to
$T_{kmax} \sim 240$ K. These components must be CNM. This is confirmed
by the detection of 21-cm line absorption against the sources 4C78.01,
4C72.01, 4C74.08, and 4C76.13 in the huge Nan\c{c}ay survey of
Crovisier, Kaz\`es, and Aubry (1978).

	Heiles maps these CNM components, so here we have another rare
opportunity to view the angular structure of CNM. His Figure 7 shows
maps of the two peaks. The maps show narrow rifts running through
otherwise large-scale and rather lumpy distributions. The rifts can only
occur if these structures are sheets. The velocities merge near one end
of the region, from which Heiles concludes that they are physically
related and could easily be the front and rear walls of an expanding
shock. 

	The high-velocity sheet (HVS) is lumpier and has
$N(HI)_{CNM,20}$ ranging up to $\sim 4$; the low-velocity sheet (LVS) is
smoother with smaller peak columns, about 2.5. Thus these sheets have
about ten times the column density of the triad region's sheets
discussed in \S \ref{triadregion}. For these sheets Heiles estimates
$L_{||} \lesssim 3.6 T_{40}/P_{2250}$ pc. On the plane of the sky,
$L_\perp \sim 50 D_{100}$ pc, so 

\begin{equation}
{L_\perp \over L_{||}} \gtrsim 14 {D_{100} P_{2250} \over T_{s,40}} \ 
\end{equation}

\noindent Heiles estimates $D_{100} \sim 5$, so $(L_\perp / L_{||}) \sim
70$; this ratio is not as spectacularly high as the triad region sheet
but is nevertheless quite impressive.

	Heiles also finds ``cloudlets'' within the sheets and summarizes
their statistical properties in his Figures 11 and 12. The areal density
is high: 815 cloudlets over 160 deg$^2$ is 5 per deg$^2$, or half this
for each sheet. The $\Delta V_{FWHM}$ histogram is narrow, $\sim 0.8$ km
s$^{-1}$ wide, and peaked at $\sim 2.0$ km s$^{-1}$, corresponding to
$T_{kmax} = 88$ K. The median column density $N(HI)_{CNM,20} \sim 0.3$,
much like the sheets in the Triad Region; this gives $L_{||} \sim 0.17
T_{s,40}/P_{2250}$ pc. The typical angular diameter is 31 arcmin, so
the cloudlets have

\begin{equation}
{L_\perp \over L_{||}} \sim 4.9 {D_{100} P_{2250} \over T_{40}} \ .
\end{equation}

\noindent With $D_{100} \sim 5$, these are also very 
sheetlike, but not so much as the sheets with which they are
associated. 

	The term ``blobby sheet'' seems to be the correct descriptive term
the large sheets in the Small Region.

\subsection{Evidence from other studies} \label{otherstudies}

	High-resolution studies of the CNM provide abundant evidence
that the CNM is not distributed in isotropic clouds.  On large scales,
the maps of HI in self-absorption by Gibson et al (2000) and Gibson
(2001) show a plethora of structures.  Commenting on their Figure 1,
Gibson et al (2000) describe it as including ``overlapping knots,
filaments, and other complex structures.'' The low-latitude gas studied
by them is quite distant, so these structures are hundreds of parsecs in
scale. Other, more localized studies of the CNM, outlined below, also
reinforce the conclusion that the CNM does not lie in isotropic clouds.

	Greisen and Liszt (1986; GL) made interferometric
high-resolution (a few arcsec) maps of the 21-cm line opacity spectra
against the extended sources 3C111, 3C161, and 3C348. They examined the
angular structure of 9 Gaussian opacity components. Obtaining
$N(HI)_{CNM}$ for their components requires assuming $T_s$; if $T_s =
40$ K, then $N(HI)_{CNM,20}$ ranges from 0.20 to 2.0. Two of the three
sources had $|b|<10^\circ$, so many components have $D_{100} \gg 1$.
Nevertheless, the fluctuation statistics of all 9 components are
similar, approximately independent of $N(HI)_{CNM}$. GL found
variations on scales $\gtrsim 30$ arcsec; they characterize those
variations as ``well-behaved'', meaning that the variations are
relatively smooth and not disorganized or chaotic. Thus, the CNM
Gaussians do not display the random polka-dot pattern expected from the
independent clouds.

	GL do see one cloud edge. Their  lowest-$N(HI)_{CNM,20}$ ($=-0.20$)
component resides towards 3C161 [$(l,b=(215.4, -8.1)$] and has $V_{LSR}=28$
km s$^{-1}$, making $D_{100} \sim 28$. Equation \ref{thetaeqn} predicts
$\theta_\perp \sim 0.08$ arcmin. In fact, they saw this component in
only two of three positions; the two positions are separated by 0.15
arcmin, and the third is 0.9 arcmin away, so the cloud is larger than we
predict by at least a factor of two. However, this is not a very serious
discrepancy; it can be fixed by adjusting $P_{2250}$ and/or $T_{s,40}$. 

	GL analyzed two components in 3C348, which is also on our source
list.  The results are given in Table \ref{glsourcepairs}.  The stronger
$V_{LSR}$ 0.5 km s$^{-1}$ component has a larger fractional variation in
profile area than the weaker one, which is contrary to our expectation
from \S \ref{drastic}. 

	Dickey (1979) analyzed pairs of opacity spectra against lobes of
double radio sources. Two, 3C348 and 3C353, have $|b|>10^\circ$. 3C353
has a strong opacity component (we find $\tau_0 = 1.2$) which shows less
than $10\%$ variation across 3.7 arcmin. He resolved 3C348's opacity
spectrum into two components, one at $V_{LSR}=0.1$ and one at 7.9 km
s$^{-1}$; the former is strong (our $\tau_0= 0.6$) and has $\lesssim
7\%$ variation and the latter is weak (our $\tau_0=0.078$) and has
$\lesssim 30\%$ variation across 1.9 arcmin. In a related study PST
compared the properties of opacity spectra against small-diameter and
large-diameter (up to a few arcmin) sources and found no statistically
significant differences. These results are a bit marginal in sensitivity
but do reinforce our conclusion. 

	Kalberla, Schwarz, and Goss (1985) used the WSRT to generate a
high-resolution HI data cube centered on 3C147, located at
$(l,b)=(161.7^\circ, 10.3^\circ)$. They were able to map the emission
produced by five Gaussian components in the opacity profile over their
field of view, which is about 30 arcmin diameter. In every case the
emission has structure on the scale of a few arcmin but is extended and
spills outside the field of view in at least one direction. This is a
very direct way to study the angular extent of the CNM and needs to be
repeated for many sources. 

\subsection{Summary: CNM component morphology must be sheetlike}

	The above comparisons of opacity profiles using both our own
data and previous literature show that the rapid angular variation in
opacity profile structure expected under the isotropic cloud model does
not occur in the sources studied. These sources are not a complete
sample and these comparisons should be extended. Nonetheless, not a
single source with HI absorption nor any HI line survey supports the
isotropic cloud model for the CNM. That is, equation \ref{thetaeqn} does
not correctly predict the scale of angular variations in CNM clouds.

	Equation \ref{thetaeqn} is based on three assumptions.
\begin{enumerate}

	\item The CNM pressure $P_{2250} \sim 1$.  This pressure is
observationally determined from observations of the CI line, which is
produced in CNM regions.  It has a significant dispersion but a
well-defined median.  This assumption is as close to an observational
fact as we get in astronomy. 

	\item The distance $D_{100} \gtrsim 1$; if a cloud becomes
arbitrary close, then it can have arbitrarily large $\theta_\perp$.  We
observe from within the Local Hot Bubble (LHB), which has a radius $\sim
50$ to 150 pc, depending on direction (Sfeir et al 1999).  The LHB is
characterized by its pervasive HIM and absence of dense clouds.  Our CNM
components cannot be produced within the LHB, so they cannot lie
arbitrarily close. 

	\item Clouds are isotropic so that $L_\perp \sim L_{||}$.  This
assumption must be wrong.  The maps for the Triad and Small Regions are
specific cases, with aspect ratios in the range 100-300, where this
assumption clearly does not apply.  Another is the recent maps of 21-cm
line self-absorption in the Galactic plane (Gibson et al 2000, Gibson
2001), which show structures with all angular scales and even a blobby
sheetlike structure extending over many degrees (because it is distant,
this means hundreds of parsecs).  \end{enumerate}

	The maps for the Triad and Small Regions are specific cases for
which the isotropic assumption does not apply, as are the low-latitude
regions mapped by Gibson et al (2000) and Gibson (2001).  We conclude
that CNM clouds are not isotropic.  To reproduce the observed situation
in which they almost always extend over much larger angles than the
$\theta_\perp$ of equation \ref{thetaeqn}, they must be sheetlike.  The
sheets are not perfectly smooth because we do see variations with
position.  They are best characterized as ``blobby sheets''. 

	In \S \ref{drastic}, we discussed the effects of beam dilution
on the derived spin temperatures and column densities for isotropic
clouds. However, clouds are not isotropic, so the effects estimated
there are greatly exaggerated. Nevertheless, these effects probably do
operate at some level because the CNM sheets are blobby. 

\section{A DIRECT COMPARISON WITH THE McKEE/OSTRIKER MODEL}

\label{momodel}

	The McKee and Ostriker (1977; MO) model of the interstellar
medium predicts each CNM component to be embedded in, and thus pressure
equilibrium with, a single WNM or WIM/WNM cloud. The warm gas acts as a
buffer between the cold, neutral, dense gas and the X-rays produced by
the Hot Ionized Medium (HIM), cosmic rays, and UV radiation from stars.
Most of the WNM envelopes should be in thermally stable equilibrium with
$T_k \sim 8000$ K.

\subsection{Method and tabular results}

\label{momethod}

	We directly investigate the applicability of the MO model to our
data by performing least-squares fits with this model directly in mind.
In contrast to our empirical method described in Paper I \S 5,
which models the WNM as a small number of Gaussians with arbitrary
centers and widths, here we model the WNM as follows: \begin{enumerate}

	\item We begin with the same CNM components as in Paper I.

	\item One portion of the WNM, the CNM-associated portion, is
represented by a set of WNM Gaussians, each WNM Gaussian having the same
central velocity as its corresponding CNM component. We assume that the
WNM components have no significant nonthermal motions, so we constrain
the width to be $T_{kmax} = T_k = 8000$ K.  We allow departures from
this constraint as described below. 

	\item The other portion of the WNM, the CNM-independent one, is
represented by one, or in a very few cases two, additional Gaussians,
with arbitrary centers and widths, that are unrelated to the CNM
components. The MO model allows this because not every WNM cloud need
have a CNM core; and our data demand it. 

	This model is surprisingly successful at fitting many of our
profiles. However, for many sources the fit is significantly improved by
allowing allow departures from assumption (2) as follows:

	\item Sometimes CNM components are spaced so closely, i.e.\ much
closer than the WNM linewidth $\Delta V_{FWHM} = 19$ km s$^{-1}$, that
the associated WNM components are degenerate.  In these cases, we use a
single WNM component for all of the closely-spaced CNM components and in
statistical discussions divide the WNM equally among the associated CNM
components; thus all of these CNM components have the same WNM column
density, which we denote by the symbol $N(HI)_{WNM;CNM}$, but of course
they have different CNM column densities $N(HI)_{CNM}$.  We always try
to pair a WNM component with each CNM one.  However, if this doesn't
work, we define a CNM component to be associated with a WNM component if
the CNM's velocity falls within the halfwidth range of the WNM Gaussian. 

	\item Sometimes it is obvious that the fit can be greatly
improved by allowing the WNM line width to vary as a free parameter.
This allows us to derive values for $T_{kmax}$ for the WNM components
that differ from 8000 K. Almost all of these have lower $T_{kmax}$, and
many of these lie in the unstable region between 500 and 5000 K.

	\item Almost always, a small change in WNM line center has
little influence on the fit quality. This is in contrast to the line
width, mentioned above. 3C274.1 is the only case where a change in WNM
line center would significantly improve the fit, but we we don't allow
the central velocity to change because we wish to keep the model as
simple as possible without generalizing it for a single exception. This
has no significant effect on the derived values of $T_{kmax}$ and no
ramifications for our discussion. 

\end{enumerate}

	Except for 3C133, 3C409, 4C13.67, and P0531+19, we excluded
sources having $|b| < 10^\circ$ from the analysis because the profiles
are too complicated. We also excluded all sources having no CNM
components. This leaves a total of 47 sources with 112 WNM components
and 142 CNM components. 82 of these 112 WNM components are associated
with the 142 CNM ones, and 30 WNM components are not associated with
CNM. 

	12 good-quality fits follow the MO model strictly in having
one-to-one paired WNM and CNM components plus perhaps an additional
CNM-independent WNM component. Including multiple CNM components per WNM
component, 38 sources have good quality fits. 3 sources, 3C142.1, 3C225b,
and 3C274.1 have poor quality fits, but no worse than for the standard
fits. The fits for 5 sources were much worse than the standard fits:
3C207, 3C315, 3C318, 3C409, and P0428+20. 3C225b also falls into this
category, but only because some narrow opacity components are not
represented by Gaussians because of its large error in the opacity
profile. 

%\clearpage
\begin{figure}[h!]
\begin{center}
\includegraphics[width=5in] {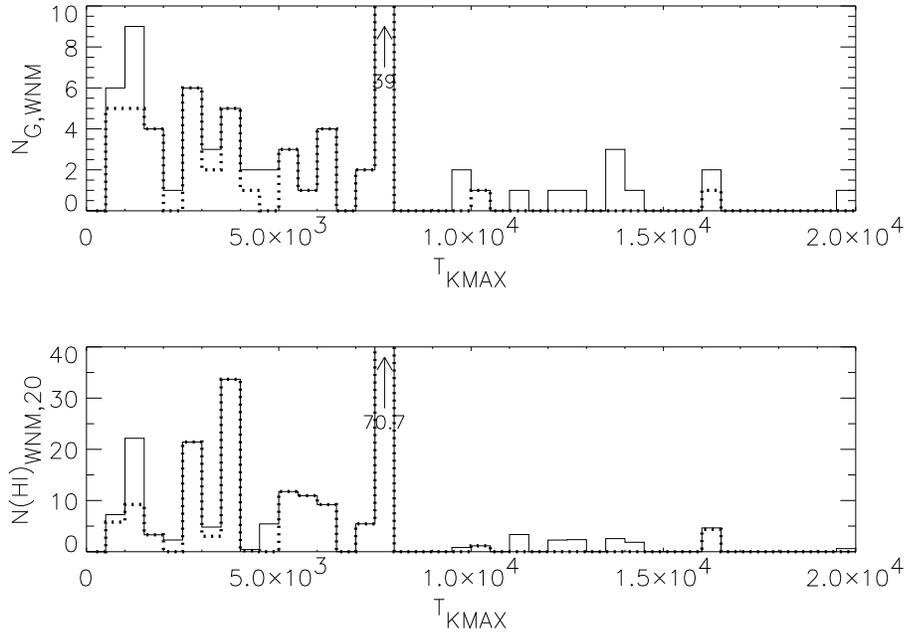} 
\end{center}

\caption{For the MO model fits, histograms of $T_{kmax}$ for the WNM
Gaussians of the MO-fit model. The dotted histogram shows CNM-associated
WNM components and the solid histogram all WNM components. The top
histogram is for number of components $N_G$ and the bottom one for
$N(HI)_{WNM}$.  Annotated arrows indicate histogram heights for all WNM
Gaussians at $T_{kmax}=8000$ K,  only one which is not associated with
CNM; it contains $N(HI)_{WNM,20}=1.0$. \label{histoplot4mo}}
\end{figure}
%\clearpage

\subsection{ Linewidths, $T_{kmax}$, CNM and WNM column densities}

\subsubsection{ Thermally unstable WNM}

	For CNM-associated WNM components, we allowed the linewidth
$T_{kmax}$ to vary as a free parameter if this would significantly
improve the fit. For many sources this adjustment was not required. When
it was required, the resulting $T_{kmax}$ was almost always less than
8000 K. The fact that most fits did not require this adjustment and that
sometimes the line width is smaller suggests that the temperature 8000 K
is, indeed, a reasonable one for much of the WNM, as predicted by MO and
subsequent theory (e.g.\ Wolfire et al 1995) and, moreover, that
nonthermal line broadening is not very important in much of the WNM (in
contrast to the CNM). 

	Figure \ref{histoplot4mo} shows histograms of $T_{kmax}$ for the
WNM components, both the CNM-associated ones (dotted histogram) and all
WNM components (solid), and both for number of components $N_G$ and for
$N(HI)_{WNM}$.  The obvious peaks at $T_{kmax}=8000$ K result from
CNM-associated WNM where we have constrained the WNM linewidth by this
temperature [see (2) in \S \ref{momethod} above].  WNM gas components
that are not associated with CNM ones were, of course, fit without a
width constraint.  Some gas [$(7\%, 6\%)$ for $(N_G, N(HI))$] has
$T_{kmax}>20000$ K and is off the histogram to the right.  Apart from
the 8000 K peak, the histogram is not dissimilar in shape to the
corresponding histograms in Figure \ref{histoplot4new}. 

	Much of the CNM-associated WNM gas, $(35\%, 40\%)$ for number of
components and column density, lies in the thermally unstable range 500
to 5000 K; the corresponding fractions for all WNM gas are $(34\%,
41\%)$. For our standard fits these fractions were $(39\%, 48\%)$ (\S
\ref{cnmwnmsummary}). These column density fractions are comparable,
which suggests that these numbers are robust. The sources analyzed here
are only those containing CNM components, which is a biased sample, so
we adopt the higher value from \S \ref{cnmwnmsummary} as our final one
and conclude that a significant fraction of all WNM, $\gtrsim 48\%$ by
mass, lies in the thermally unstable range 500 to 5000 K.

%\clearpage
\begin{figure}[h!]
\begin{center}
\includegraphics[width=3.5in] {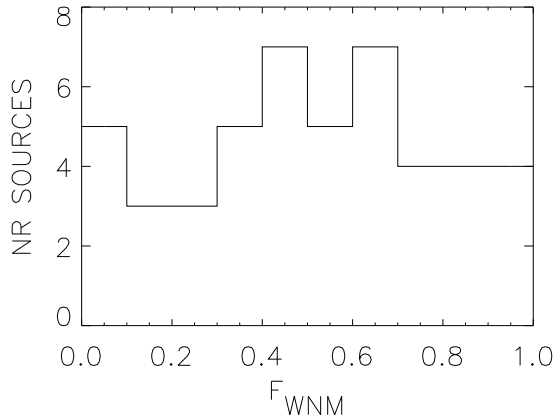} 
\end{center}

\caption{For the MO model fits, histogram of $F_{WNM}$, the
column-density fraction of all WNM along a line of sight that has
$T_{kmax}$ in the thermally unstable range 500 to 5000 K. 
\label{histomapmo}} \end{figure}

%\clearpage

\begin{figure}[h!]
\begin{center}
\includegraphics[width=3.5in] {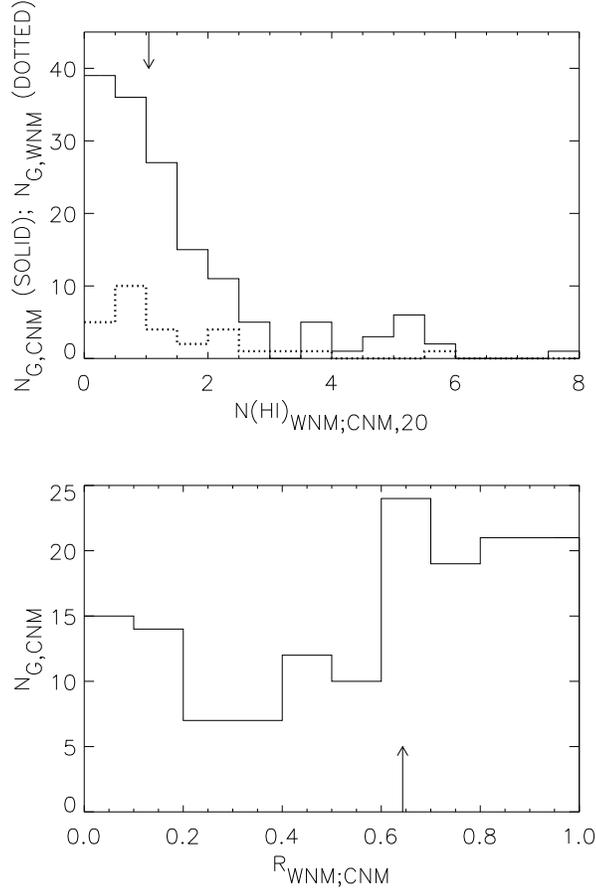} 
\end{center}

\caption{For the MO model fits. {\it Top}, the solid line is the
histogram of $N(HI)_{WNM;CNM}$, the total WNM column
density associated with each CNM component; the dotted line is the
histogram of WNM {\it not} associated with CNM (for which the horizontal
axis should be labeled $N(HI)_{WNM,20}$). $N_{G,CNM}$ is the number of
CNM Gaussian components. {\it Bottom}, histogram of $R_{WNM;CNM}$ for
each CNM component. See \S \ref{wnmfractioncnm} and \S
\ref{wnmratiocnm}.  \label{histoplotmo1}} \end{figure}
%\clearpage

	Consider the fraction $F_{WNM}$ of all WNM gas along a line of
sight that lies in the thermally unstable range 500 to 5000 K.  Figure
\ref{histomapmo} shows a histogram of this ratio.  The distribution is
roughly flat, with no preference for any particular ratio.  Figure
\ref{mapnhratio} (bottom) shows a map of this quantity; the highest
values seem to cluster in the Taurus-Perseus and NPS regions.  This
suggests a correlation between $F_{WNM}$ and $R(HI)_{CNM}$.  The
correlation coefficient is 0.29, but the scatter plot is not very
impressive to the eye.  We conclude that thermally unstable gas is
common and not closely related to other physical parameters. 

\subsubsection{Fraction of WNM gas}

\label{wnmfractioncnm}

	The MO model predicts the WNM column density associated with a
CNM core, which we define as $N(HI)_{WNM;CNM}$; this notation is meant
to mirror that of conventional statistics, i.e.\ the WNM column density
given a certain CNM one.  It is calculated as described in (2) and (4)
of \S \ref{momethod}.  Figure \ref{histoplotmo1} shows two histograms of
$N(HI)_{WNM;CNM}$.  The bottom panel shows the fraction

\begin{equation} \label{wnmratiocnm}
R_{WNM;CNM} = {N(HI)_{WNM;CNM} \over N(HI)_{WNM;CNM} + N(HI)_{CNM} } \ ,
\end{equation}

\noindent which is, for a particular CNM component, the ratio of its
associated WNM column density to total CNM-associated column density
(both WNM and CNM).

	The MO model predicts that every CNM cloud is enveloped in WIM
and that some, $\sim 1/3$, are also enveloped in WNM. MO Figure 1 shows
a typical small cloud, which has  with typical total column densities
through the diameter $N(HI)_{WNM,20} \sim 0.03$ and $N(HI)_{CNM,20} \sim
1.3$. There is an additional WNM contribution from the partially-ionized
WIM; all of this gives $R_{WNM;CNM} \sim 0.04$. MO's numbers apply at
$z=0$ and are predicted to increase with $|z|$. These numbers are very
rough, but do not agree well with the observational data in Figure
\ref{histoplotmo1}.

	This disagreement is simply a different expression of the large
WNM fraction in the ISM, which disagrees with the MO model.  The MO
model predicts a much smaller ratio of WNM to CNM column densities than
we observe, whether or not we fit our observations in terms of the MO
model or with the empirically oriented method of Paper I. Overall, MO
predict that about $4\%$ of the diffuse interstellar gas should be WNM
(this includes the WIM-associated HI). Yet here and in \S
\ref{coldenstatistics} the ratio is much larger. The overall ratio of
all WNM to total for this MO-oriented fit, whether or not the WNM is
associated with CNM, is $R(HI)_{WNM}=  1 - R(HI)_{CNM} = 0.57$, which is
more than ten times what MO predict. For the entirety of sources treated
using the empirical method of Paper I, $R(HI)_{WNM} = 0.61$. This latter
number is a bit higher and is better because the MO-model sample is
restricted and biased. 

\subsection{ Summary of comparison with MO}

	The data compare with the MO model in the following respects:
\begin{enumerate}

	\item Modeling WNM profiles as envelopes having the same
velocity as their associated CNM Gaussians works very well for most
sight lines, but for some it works poorly. 

	\item The WNM column densities in the CNM envelopes are far
larger than predicted. 

	\item Overall, the WNM constitutes about $61\%$ of the total HI,
more than ten times the predicted fraction. However, our observed number
referes to all $|z|$, while MO's refers to $z=0$.

	\item At least $\sim 48\%$ of the WNM is thermally unstable. MO
would allow only a small fraction, that portion of the gas that is
transiting from one phase to another.

\end{enumerate}

\section{TWO DESCRIPTIVE MODELS BASED ON OBSERVATIONS}

\label{descmodel}

\subsection{The raisin-pudding model: not applicable}

	First we discuss the CNM statistics in terms of the often-used
conceptual model of randomly distributed isotropic clouds embedded in a
WNM substrate, which we term the ``raisin-pudding'' model. This model
is popular and deserves to be addressed, despite the fact that we have 
shown in \S \ref{againstraisin} that CNM clouds are not isotropic. 

	For purposes of discussion we will suppose that the WNM has
typical $T_{k}=4000$ K; if $T_k = T_{kmax}$ this is not an unreasonable
discussion value, and is convenient because it is 100 times our adopted
CNM temperature. If the WNM has the same pressure as the CNM, then its
volume density is 100 times smaller. Our $N(HI)$ histograms show that
the WNM column density $N(HI)_{WNM}$ is typically larger by a factor
$\sim 1.5$ than $N(HI)_{CNM}$. This makes the typical ratios
$L_{||,WNM}/L_{||,CNM}$ and $\theta_{\perp,WNM}/\theta_{\perp,CNM} =
150$ (for definitions, see \S \ref{isotropic}). 

	Suppose that both the CNM and WNM consist of isotropic clouds of
diameter $L_{||,CNM}$ and $L_{||,WNM}$, respectively.  The CNM Gaussian
component clouds are much smaller than the WNM ones, so we imagine that
the CNM clouds are embedded in a single WNM Gaussian component cloud of
diameter $L_{||,WNM}$---like raisins in a giant pudding.  The number of
CNM components that should be observed along a typical line of sight is
$N_{CNM} \sim (L_{||,WNM}/S_{||,CNM})$, where $S_{||,CNM}$ is the mean
free path for a line of sight intersecting the CNM clouds.  The mean
free path is

\begin{equation}
S_{||,CNM}= {1 \over \nu \sigma_{CNM}}
\end{equation}

\noindent where $\nu$ is the number of CNM clouds per unit volume and
$\sigma_{CNM}$ is the effective cross section of a cloud; for a
spherical cloud, the effective size is the diameter plus the diameter
$L_O$ of the sampling beam (which can be the radio source for absorption
and the telescope beam for emission), so we can write

\begin{equation}
\nu = {4N_{CNM} \over \pi L_{||,WNM} \left( L_{||,CNM} + L_O \right)^2 }
\end{equation}

\noindent and the total number of CNM clouds residing within the WNM
cloud is

\begin{equation}
{\cal N} = {2N_{CNM} \over 3} { (L_{||,WNM} / L_{||,CNM})^2 \over
   \left[ 1 + (L_O / L_{||,CNM}) \right]^2 }
\end{equation}

	This number is enormous. For $N_{CNM}>1$ and
$(L_{||,WNM}/L_{||,CNM}) = 150$, it exceeds $2 \times 10^4$.

\subsection{The CNM clumpy sheet model: better}

	The ISM contains more WNM than CNM. The are many lines of sight 
that contain WNM but no CNM. The WNM is extended over path lengths of
100 pc or more. This does not require or even suggest the MO concept in
which each CNM cloud has a separate, independent WNM envelope; if this
were the case in fact, then with just a few CNM components their
associated WNM envelopes would merge into a single WNM cloud. This
points towards a model in which the WNM occupies large volumes and CNM
components lie inside. 

	From \S \ref{againstraisin} we find that the CNM components are
sheetlike.  From \S \ref{otherstudies} we find that the CNM sometimes
appears as elongated filaments. A continuous, wrinkled sheet can look
like a filament where the sheet happens to lie tangent to the line of
sight (Hester 1987). Also, a thin ribbon can also be perceived as a
filament. 

	If all CNM sheets had the same column density thickness,
then the observed $N(HI)_{CNM}$ would increase as the sheets become more
tangent to the line of sight.  With a random distribution orientation
large tilt angles are preferred, so the histogram of $N(HI)_{CNM}$
should increase markedly towards large values.  Figure
\ref{histoplotnhgaussians} shows that it doesn't.  This means that the
intrinsic column density thickness has a wide dispersion: some sheets
are thin, some are thick.  In two regions for which we are fortunate
enough to have CNM maps, the CNM is distributed in huge blobby sheets of
thickness $\sim 0.11$ and $\lesssim 3.6$ pc, with length-to-thickness
aspect ratios $\sim 280$ and $\sim 70$. 

	If these characteristics are general, then the CNM seems to be
organized into a small number of large, thin structures.  In contrast to
the raisin pudding model, in which the CNM blobs are spherical and
randomly distributed with $\gtrsim 2 \times 10^4$ CNM cloudlets within a
WNM cloud, there are only a few such sheets. The sheets probably contain
lots of blobs; in the Small Region the density $\sim 2.5$ cloudlets per
deg$^2$, or 1 cloudlet per 30 pc$^2$. This is conceptually a much
different morphological arrangement than the raisin pudding model.

	The arrangement in large sheets is consistent with ideas that
the CNM forms from large-scale shocks produced, for example, by
supernovae or large-scale vertical shocks (Walters \& Cox 2001). The
Small Region's sheets merge in velocity and are suggestive of what we
expect from an expanding shell, and were originally so interpreted. 

	However, invoking a shock for the Triad Region has its
difficulties. First we remark on a favorable situation for the shock
interpretation, namely the cold temperature ($T_s \sim 20$ K), which
suggests an absence of grain heating, and the grains could have been
destroyed by the shock. The sheet's $V_{LSR}$ is small, suggesting a
shocked shell that has suffered substantial deceleration. But the
velocity fluctuations are also small, which is unexpected because the
deceleration should occur in a clumpy medium, producing large velocity
fluctuations. In particular, we would expect large fluctuations for a
sheet with small column density, which is the case here ($N(HI)_{CNM,20}
\sim 0.3$).

	This clumpy sheet model must be considered provisional because
it is based on extrapolating mapping results from only two regions to
the entire ISM. We desperately need CNM maps for more regions. New maps
of self-absorption of the 21-cm line in the Galactic plane are being
produced by the current interferometric HI surveys, e.g.\ Gibson et al
2000. Maps away from the Galactic plane are also important because they
allow detailed study of regions with less confusion caused
foreground/background gas.

\section{SUMMARY} \label{summary}

	Paper I discusses the observational and data reduction
techniques. In particular, it devotes considerable attention to the
Gaussian fitting process, which is subjective and nonunique. Concerned
readers should see \S 5 of that paper.

	The present paper treats the astronomically oriented
implications of the Gaussian components from Paper I and includes the
following topics: \begin{enumerate} 

\item \S \ref{tempdist} discusses the statistics of the Gaussian
components. It shows that the CNM and WNM are not only observationally
distinct, but also physically distinct. The median column density per
CNM Gaussian component is about $0.5 \times 10^{20}$ cm$^{-2}$, and per
WNM component about $1.3 \times 10^{20}$ cm$^{-2}$ (Table
\ref{medianmeannh}). 

	The CNM temperature histogram peaks near $T_s = 40$ K (Figure
\ref{histoplot4new}), about half the temperature obtained by previous
workers. Its median by components is 48 K and, weighted for $N(HI)$, 70
K. CNM temperatures range down to $\sim 15$ K, which can be attained
only if grain heating is not operative.  CNM temperatures appear to be
smaller than those derived from UV absorption line observations of
H$_2$, but the comparison means little because H$_2$ temperatures refer
to all velocity components and all phases along the line of sight. 

	A significant fraction of the WNM, $\gtrsim 48\%$, lies in the
thermally unstable range $T_k = 500$ to 5000 K.

\item \S \ref{coldenstatistics} summarizes the statistics of WNM and CNM
column densities for entire lines of sight instead of individual
Gaussian components. There are many lines of sight having no CNM; these
form a distinct class and are confined to particular areas of the sky.
Column densities depart very markedly from those expected from a
plane-parallel distribution. $61\%$ of the HI we observed is WNM; at
$z=0$, it fills $\sim 50\%$ of the volume, but this number is {\it very
rough}. In \S \ref{volfill} we show that this is in reasonably good
agreement with MO, when the WIM-associated HI is included.

Figure \ref{mapnhrawratio2} shows the factor $R_{raw}$ by which $N(HI)$
calculated from the optically thin approximation (i.e.\ from the line
profile area) underestimates the true $N(HI)$; this can be significant
even at high Galactic latitudes.  

\item \S \ref{vlsrstatistics} shows that the component velocities that
we observe are not significantly affected by Galactic rotation. The
column-density weighted rms velocities are about 7 and 11 km s$^{-1}$
for the CNM and WNM Gaussian components, respectively. 

\item \S \ref{correlations} uses Principal Components Analysis, together
with a form of least squares fitting that accounts for errors in both
the independent and dependent parameters, to discuss the relationships
among the four CNM Gaussian parameters. The spin temperature $T_s$ and
column density $N(HI)$ are, approximately, the two most important
eigenvectors; as such, they are convenient, physically meaningful
primary parameters for describing CNM clouds. 

	The Mach number of internal macroscopic motions for CNM clouds
is typically $\sim 3$, but there are wide variations and a weak increase
with $T_s$. Most CNM clouds are strongly supersonic. We discuss the
historical $\tau_0$-$T_s$ relationship in some detail and show that it
has little physical meaning. 

\item \S \ref{pequality} discusses the possible effect of angular
resolution on the relationships among observed CNM parameters. These
effects are important if CNM clouds are isotropic. However, 
\S \ref{againstraisin} shows that CNM clouds are definitely not
isotropic. CNM features are sometimes large sheets with aspect ratios
measured in the hundreds. These sheets contain blobs, which themselves
are sheetlike but with much smaller aspect ratios.

\item \S \ref{momodel} directly compares our data with the McKee/Ostriker
model by re-reducing all Gaussian components in terms of that model,
i.e.\ with each CNM cloud having an associated WNM envelope. This
fitting scheme works very well for many sources, but not for all.
The MO model greatly underpredicts the WNM abundance and, also, the
fraction of WNM that is thermally unstable. 

\item In \S \ref{descmodel} we argue that there is so much WNM that CNM
clouds probably don't have individual WNM halos, but rather that many
CNM clouds exist within a common WNM halo. We discard the raisin pudding
model as a commonly envisioned descriptive model and replace it by the
blobby sheet model, in which the CNM consists of sheetlike structures
with sheetlike blobs or cloudlets embedded within. Each WNM cloud
probably contains a few CNM large sheets.

\item \S \ref{againstraisin} uses our knowledge of the CNM pressure to
derive the morphological shape of CNM structures: they are sheetlike. In
two regions of the sky the CNM is organized into large sheets with
length-to-thickness aspect ratios $\sim 280$ and 70; the latter is
permeated by small sheetlike structures. 

\item In the following section we provide comments on the importance of
the WNM for understanding not only the ISM but also the full range of
its energy sources.

\end{enumerate}

\section{THE WNM: KEY TO THE UNIVERSE}

\label{commentary}

	From the theoretical standpoint, Wolfire et al (1995; WHMT) show
that the temperature of the CNM is well constrained: if the density is
large enough, the time scale for equilibrium is short and the
equilibrium temperature is well defined. Their predicted temperature is
close to the peak in our CNM histogram, so our data are very consistent
with their results. Anomalies with colder temperatures such as the Triad
Region's sheets can be achieved if grain heating does not operate; these
regions are fascinating, but not very common.

	The WNM is another matter. Theoretically, the temperature is
well constrained, but the time scale for equilibrium is not short.
Moreover, there are formally forbidden ranges in density and temperature
because of the thermal instability. In fact the thermal time scales are
long enough that a sufficiently chaotic medium might never reach thermal
equilibrium. Theories like WHMT's that discuss only the thermal
equilibrium microphysics cannot easily deal with these matters.

	Our finding that much of the WNM lies in the thermally unstable
range 500 to 5000 K strongly implies that thermal equilibrium does not,
in fact, obtain for much of the WNM.  Moreover, the WNM seems to have
significant ionization, with a mean ionization fraction possibly as high
as 0.2 but with large fluctuations (Heiles 2001a). It strikes us that the
amount,  thermal state, and ionization state of the WNM are sensitive
indicators of the conflicting effects of dynamical (macrophysical) and
atomic (microphysical) processes, both of which heat and cool the gas.
In addition, microphysical processes heat by ionizing the gas, while
macrophysical ones usually do not. 

\subsection{Microphysical processes}

	Microphysical processes include the ones treated by WHMT, which
rely on well-known radiation energy densities. However, these are not
necessarily so well-known as we would wish. Consider, for example, the
production of the Warm Ionized Medium (WIM) by ionizing photons.
Classically, we expect ionizing photons to be strictly limited to their
Str\"omgren spheres; in fact, however, the photons can diffuse out to
large distances and produce the WIM, which produces pulsar dispersion
and diffuse H$\alpha$ emission. The diffusion efficiency is only
partially understood (Miller \& Cox 1993; Dove \& Shull 1993). This
shows that we do not completely understand photon propagation in the
ISM.

	We wish to mention two additional microphysical processes that
might be underappreciated and add significant heating. Both of these act
preferentially on low-density gas and thus affect the WNM more than the
CNM. The first process is low-energy cosmic rays, whose energy density
cannot be measured directly because they are excluded from the Solar
System.  Geballe et al (1999) observe H$_3^+$ to be much more abundant
than predicted in diffuse clouds; a probable reason is a considerable
excess of low-energy cosmic rays over the current standard value. Such
cosmic rays ionize and heat the ISM.

	The second is X-rays from soft gamma-ray repeaters. Consider the
specific example of the famous 27 Aug 1998 event of SGR 1900+14, which
was the most powerful of many bursts produced by an object $\sim 6$ kpc
distant (see Feroci et al 2001 for a review). This particular burst
produced enough X rays to ionize the nighttime Earth's atmosphere to the
extent normally found in daytime. This, in turn, required X rays whose
energies are so large that they are of little interest for ISM heating
(because the interaction cross sections are small). However, it strikes
us as unlikely that the intrinsic X-ray spectrum cuts off at low
energies. Rather, the lower-energy X rays are easily absorbed by the
ISM. Bursts from the ensemble of gamma-ray repeaters in a galaxy might
be a significant energy source for heating the WNM. If so, the limited
lifetime of soft gamma ray repeaters would probably produce conditions
mimicking  time-dependent models of the ISM such as that of Gerola,
Kafatos, \& McCray (1974). 

\subsection{Macrophysical processes}

	There exist several dynamical processes that can heat the ISM.
These, like the microphysical ones mentioned above, preferentially heat
the WNM over the CNM. These processes include hydromagnetic wave heating
(Ferriere, Zweibel, \& Shull 1988), MHD turbulence (Mintner \& Spangler
1997), magnetic reconnection (Vishniac \& Lazarian 1999), scattered
shocks (acoustic waves or ``thunder''; Ikeuchi \& Spitzer 1984),
turbulence (e.g.\ Gazol et al 2001), and turbulent mixing layers at the
boundaries of neutral clouds (Slavin, Shull, \& Begelman 1993).  When we
think of shocks we usually think of supernovae. However, shocks are
produced by other methods on both small and large scales. Examples at 
small scales include ejecta from newly forming stars, HII regions, and
cloud collisions. At large scales we have Galactic dynamics and
gravitation of large clouds (Wada \& Norman 1999, 2001; Walters \& Cox
2001). 

	Some of the above-quoted references calculate distribution
functions of gas temperature and density. They find thermally unstable
gas and conclude that macroscopic dynamical processes overshadow the
microscopic ones in determining gas temperature. These macrophysical
processes are hard to calculate because they depend indirectly on
coupling to many forms of energy input.

\subsection{Commentary}

	The WNM is the key to the Universe because the amount,
temperature, and ionization state of the WNM depend on many processes.
Most of the processes we have mentioned depend on energy sources that
cannot be characterized without a {\it global} understanding of {\it
many} different types of objects, and most of these we know very little
about. When global ISM models are successful in predicting the observed
WNM properties, including the amount, thermal state, and ionization
state, then we will have made a significant step forward in
understanding many aspects of not only the interstellar medium but also 
all of its associated energy sources. These include many objects of
general interest in the Galaxy such as, for example, the Galactic
dynamo, spiral density wave shocks, supernovae, and soft gamma-ray
repeaters.

\acknowledgements

	We thank Leo Blitz, Tom Dame, James Graham, Dave Hollenbach, Ed
Jenkins, Chris McKee, Yaron Sheffer, Mike Shull, Phil Solomon, Patricia
Vader, and Mark Wolfire for helpful discussions.  CH is indebted to the
UC Berkeley Astronomy Department for providing the freedom and time to
construct and teach a course in numerical data analysis, from which
experience some of the current data analysis greatly benefited. This
work was supported in part by NSF grants AST-9530590, AST-0097417,
AST-9988341, and by the NAIC.

%\clearpage

\clearpage

\begin{deluxetable}{rrrrrrrrrr}
\tabletypesize{\footnotesize}
\tablewidth{0pc}
\tablecaption{Source list \label{sourcelist}}
\tablehead{
\colhead{Source}      &
\colhead{ RA$_{1950}$ } &
\colhead{  DEC$_{1950}$ } &
\colhead{$ l $} &
\colhead{$ b $} &
\colhead{ FLUX (Jy) } &
\colhead{$ N(HI)_{WNM} $} &
\colhead{$ N(HI)_{CNM} $} &
\colhead{$ N(HI)_{tot} $} &
}
%\clearpage
%\include{tab1}
\startdata
      3C18 &  00 38 14 &  09 46 55 &  118.62 &  -52.73 & $  5.02 \pm 0.07 $ &   0.75 &   5.23 &   5.98\\
    3C33-1 &  01 06 12 &  13 02 31 &  129.44 &  -49.34 & $  8.70 \pm 0.00 $ &   0.86 &   1.95 &   2.80\\
      3C33 &  01 06 14 &  13 03 36 &  129.45 &  -49.32 & $  8.84 \pm 0.14 $ &   1.14 &   1.64 &   2.78\\
    3C33-2 &  01 06 17 &  13 06 21 &  129.46 &  -49.28 & $  3.75 \pm 0.00 $ &   1.02 &   1.90 &   2.92\\
      3C64 &  02 19 19 &  08 13 18 &  157.77 &  -48.20 & $  1.78 \pm 0.00 $ &   3.42 &   2.91 &   6.34\\
    3C75-1 &  02 55 00 &  05 51 49 &  170.22 &  -44.91 & $  2.83 \pm 0.00 $ &   5.92 &   2.05 &   7.97\\
      3C75 &  02 55 05 &  05 50 43 &  170.26 &  -44.91 & $  3.93 \pm 0.04 $ &   5.40 &   2.48 &   7.88\\
    3C75-2 &  02 55 09 &  05 49 14 &  170.30 &  -44.92 & $  2.44 \pm 0.00 $ &   6.09 &   2.14 &   8.22\\
      3C78 &  03 05 49 &  03 55 13 &  174.86 &  -44.51 & $  7.22 \pm 0.07 $ &   4.25 &   5.82 &  10.07\\
      3C79 &  03 07 11 &  16 54 35 &  164.15 &  -34.46 & $  4.25 \pm 0.00 $ &   2.46 &   6.91 &   9.36\\
     CTA21 &  03 16 09 &  16 17 39 &  166.64 &  -33.60 & $  8.22 \pm 0.00 $ &   6.43 &   3.13 &   9.56\\
  P0320+05 &  03 20 41 &  05 23 33 &  176.98 &  -40.84 & $  2.67 \pm 0.00 $ &   6.15 &   5.04 &  11.19\\
   NRAO140 &  03 33 22 &  32 08 36 &  159.00 &  -18.76 & $  2.62 \pm 0.00 $ &  16.06 &  13.42 &  29.49\\
    3C93.1 &  03 45 35 &  33 44 05 &  160.04 &  -15.91 & $  2.10 \pm 0.00 $ &   8.83 &   3.50 &  12.33\\
  P0347+05 &  03 47 07 &  05 42 33 &  182.27 &  -35.73 & $  3.06 \pm 0.00 $ &   6.18 &   7.26 &  13.44\\
    3C98-1 &  03 56 07 &  10 15 22 &  179.86 &  -31.09 & $  4.00 \pm 0.11 $ &   4.38 &   5.99 &  10.37\\
      3C98 &  03 56 11 &  10 17 40 &  179.84 &  -31.05 & $  6.18 \pm 0.00 $ &   4.92 &   6.10 &  11.02\\
    3C98-2 &  03 56 14 &  10 18 59 &  179.83 &  -31.02 & $  6.21 \pm 0.00 $ &   5.19 &   5.05 &  10.25\\
     3C105 &  04 04 44 &  03 33 25 &  187.63 &  -33.61 & $  3.74 \pm 0.32 $ &   3.26 &  11.42 &  14.68\\
     3C109 &  04 10 55 &  11 04 35 &  181.83 &  -27.78 & $  3.46 \pm 0.08 $ &   5.31 &  15.52 &  20.82\\
  P0428+20 &  04 28 06 &  20 31 11 &  176.81 &  -18.56 & $  3.66 \pm 0.00 $ &  17.00 &   6.90 &  23.90\\
     3C120 &  04 30 31 &  05 14 58 &  190.37 &  -27.40 & $  5.71 \pm 0.03 $ &   8.04 &   7.90 &  15.93\\
     3C123 &  04 33 55 &  29 34 13 &  170.58 &  -11.66 & $ 53.55 \pm 2.11 $ &  19.75 &   7.62 &  27.37\\
     3C131 &  04 50 10 &  31 24 31 &  171.44 &   -7.80 & $  2.99 \pm 0.13 $ &  17.27 &  11.28 &  28.55\\
     3C132 &  04 53 42 &  22 44 41 &  178.86 &  -12.52 & $  3.83 \pm 0.03 $ &  16.16 &   7.66 &  23.81\\
     3C133 &  04 59 54 &  25 12 11 &  177.73 &   -9.91 & $  5.93 \pm 0.04 $ &  19.15 &   9.35 &  28.50\\
     3C138 &  05 18 16 &  16 35 25 &  187.41 &  -11.34 & $  7.31 \pm 0.12 $ &   9.16 &  10.70 &  19.85\\
   3C141.0 &  05 23 27 &  32 47 35 &  174.53 &   -1.31 & $  2.01 \pm 0.04 $ &  29.05 &  23.64 &  52.69\\
  T0526+24 &  05 26 05 &  24 58 30 &  181.36 &   -5.19 & $  1.13 \pm 0.00 $ &  26.33 &  70.53 &  96.86\\
   3C142.1 &  05 28 48 &  06 28 16 &  197.62 &  -14.51 & $  3.13 \pm 0.00 $ &  13.85 &   8.11 &  21.96\\
  P0531+19 &  05 31 47 &  19 25 17 &  186.76 &   -7.11 & $  6.90 \pm 0.12 $ &  14.30 &   9.54 &  23.84\\
  T0556+19 &  05 56 58 &  19 08 45 &  190.09 &   -2.17 & $  0.97 \pm 0.00 $ &  53.63 &   0.00 &  53.63\\
   4C22.12 &  06 00 50 &  22 00 54 &  188.05 &    0.05 & $  2.16 \pm 0.05 $ &  31.58 &  53.65 &  85.23\\
     3C154 &  06 10 42 &  26 05 27 &  185.59 &    4.00 & $  5.39 \pm 0.02 $ &  26.72 &   8.84 &  35.57\\
  T0629+10 &  06 29 29 &  10 24 16 &  201.53 &    0.51 & $  2.60 \pm 0.05 $ &  22.23 &  37.02 &  59.25\\
     3C167 &  06 42 36 &  05 34 48 &  207.31 &    1.15 & $  1.72 \pm 0.01 $ &  19.39 &  30.85 &  50.24\\
   3C172.0 &  06 59 04 &  25 18 06 &  191.20 &   13.41 & $  2.56 \pm 0.00 $ &   7.31 &   0.40 &   7.71\\
 DW0742+10 &  07 42 48 &  10 18 33 &  209.80 &   16.59 & $  3.47 \pm 0.00 $ &   2.43 &   0.00 &   2.43\\
   3C190.0 &  07 58 45 &  14 23 02 &  207.62 &   21.84 & $  2.41 \pm 0.00 $ &   2.82 &   0.00 &   2.82\\
     3C192 &  08 02 35 &  24 18 34 &  197.91 &   26.41 & $  4.41 \pm 0.02 $ &   3.50 &   0.47 &   3.97\\
  P0820+22 &  08 20 28 &  22 32 46 &  201.36 &   29.68 & $  2.17 \pm 0.00 $ &   4.23 &   0.00 &   4.23\\
     3C207 &  08 38 01 &  13 23 06 &  212.97 &   30.14 & $  2.48 \pm 0.05 $ &   4.34 &   0.91 &   5.24\\
   3C208.0 &  08 50 23 &  14 04 16 &  213.66 &   33.16 & $  2.51 \pm 0.03 $ &   2.99 &   0.00 &   2.99\\
   3C208.1 &  08 51 54 &  14 17 16 &  213.60 &   33.58 & $  2.24 \pm 0.03 $ &   2.76 &   0.00 &   2.76\\
     3C223 &  09 36 50 &  36 07 41 &  188.40 &   48.66 & $  1.47 \pm 0.00 $ &   0.98 &   0.00 &   0.98\\
    3C225a &  09 39 25 &  14 05 36 &  219.87 &   44.02 & $  1.34 \pm 0.01 $ &   1.89 &   1.51 &   3.40\\
    3C225b &  09 39 32 &  13 59 30 &  220.01 &   44.01 & $  3.78 \pm 0.03 $ &   2.42 &   0.86 &   3.28\\
   3C228.0 &  09 47 27 &  14 34 00 &  220.40 &   45.99 & $  3.48 \pm 0.07 $ &   2.24 &   0.37 &   2.61\\
     3C234 &  09 58 56 &  29 01 40 &  200.21 &   52.70 & $  4.64 \pm 0.00 $ &   1.61 &   0.00 &   1.61\\
     3C236 &  10 03 05 &  35 08 49 &  190.06 &   53.98 & $  2.66 \pm 0.00 $ &   1.20 &   0.00 &   1.20\\
     3C237 &  10 05 22 &  07 44 58 &  232.12 &   46.63 & $  7.66 \pm 0.07 $ &   0.65 &   1.55 &   2.20\\
     3C245 &  10 40 06 &  12 19 15 &  233.12 &   56.30 & $  3.12 \pm 0.08 $ &   1.55 &   0.48 &   2.04\\
  P1055+20 &  10 55 37 &  20 08 02 &  222.51 &   63.13 & $  2.64 \pm 0.29 $ &   1.20 &   0.36 &   1.56\\
  P1117+14 &  11 17 51 &  14 37 22 &  239.45 &   65.26 & $  2.39 \pm 0.00 $ &   1.57 &   0.00 &   1.57\\
   3C263.1 &  11 40 49 &  22 23 37 &  227.20 &   73.77 & $  3.14 \pm 0.00 $ &   1.69 &   0.00 &   1.69\\
   3C264.0 &  11 42 32 &  19 53 56 &  235.70 &   73.05 & $  4.22 \pm 0.00 $ &   1.73 &   0.00 &   1.73\\
   3C267.0 &  11 47 22 &  13 04 00 &  254.81 &   69.68 & $  2.27 \pm 0.00 $ &   2.32 &   0.00 &   2.32\\
   3C272.1 &  12 22 32 &  13 09 40 &  278.21 &   74.48 & $  5.57 \pm 0.00 $ &   2.04 &   0.36 &   2.40\\
     3C273 &  12 26 32 &  02 19 39 &  289.95 &   64.36 & $ 56.13 \pm 1.12 $ &   1.43 &   0.50 &   1.93\\
   3C274.1 &  12 32 57 &  21 37 06 &  269.87 &   83.16 & $  2.19 \pm 0.02 $ &   2.06 &   0.30 &   2.35\\
   4C07.32 &  13 13 46 &  07 18 18 &  320.42 &   69.07 & $  1.55 \pm 0.00 $ &   1.79 &   0.32 &   2.11\\
   4C32.44 &  13 23 58 &  32 09 53 &   67.24 &   81.04 & $  4.47 \pm 0.05 $ &   0.91 &   0.14 &   1.05\\
     3C286 &  13 28 49 &  30 46 02 &   56.53 &   80.67 & $ 18.36 \pm 0.00 $ &   2.05 &   0.00 &   2.05\\
     3C293 &  13 50 02 &  31 41 43 &   54.61 &   76.06 & $  4.50 \pm 0.00 $ &   1.29 &   0.00 &   1.29\\
   4C19.44 &  13 54 42 &  19 33 44 &    8.99 &   73.04 & $  2.52 \pm 0.11 $ &   2.66 &   0.00 &   2.66\\
   4C20.33 &  14 22 37 &  20 14 01 &   19.54 &   67.46 & $  1.89 \pm 0.01 $ &   2.15 &   0.53 &   2.68\\
     3C310 &  15 02 48 &  26 12 36 &   38.50 &   60.21 & $  5.12 \pm 0.04 $ &   2.60 &   1.11 &   3.71\\
     3C315 &  15 11 31 &  26 18 37 &   39.36 &   58.30 & $  4.49 \pm 0.03 $ &   2.54 &   2.22 &   4.76\\
     3C318 &  15 17 50 &  20 26 54 &   29.64 &   55.42 & $  2.90 \pm 0.02 $ &   3.01 &   1.74 &   4.75\\
     3C333 &  16 15 05 &  21 14 51 &   37.30 &   42.97 & $  1.89 \pm 0.01 $ &   3.99 &   1.10 &   5.09\\
     3C348 &  16 48 40 &  05 04 28 &   23.05 &   28.95 & $ 46.11 \pm 0.75 $ &   4.15 &   1.55 &   5.70\\
     3C353 &  17 17 54 &  00-55 55 &   21.20 &   19.64 & $ 48.76 \pm 1.70 $ &   3.84 &   7.00 &  10.85\\
   4C13.65 &  17 56 13 &  13 28 42 &   39.31 &   17.72 & $  2.40 \pm 0.06 $ &   7.72 &   1.46 &   9.18\\
   4C13.67 &  18 35 12 &  13 28 03 &   43.50 &    9.15 & $  1.69 \pm 0.01 $ &  12.76 &   3.96 &  16.72\\
     3C409 &  20 12 18 &  23 25 42 &   63.40 &   -6.12 & $ 17.08 \pm 0.15 $ &  19.73 &   6.06 &  25.79\\
     3C410 &  20 18 03 &  29 32 35 &   69.21 &   -3.77 & $ 10.06 \pm 0.00 $ &  32.78 &  15.44 &  48.22\\
     3C433 &  21 21 30 &  24 51 17 &   74.48 &  -17.69 & $ 13.22 \pm 0.15 $ &   5.06 &   2.83 &   7.89\\
   3C454.0 &  22 49 07 &  18 32 44 &   87.35 &  -35.65 & $  2.29 \pm 0.03 $ &   4.13 &   1.24 &   5.37\\
   3C454.3 &  22 51 29 &  15 52 56 &   86.11 &  -38.18 & $ 17.22 \pm 0.38 $ &   4.80 &   1.72 &   6.53\\
\enddata
\tablecomments{ Flux is in Jy and includes the contribution from
all extended components; HI column densities are in units of
$10^{20}$ cm$^{-3}$. } 
\tablecomments{ Some sources having $|b|<10^\circ$ have very complicated
HI profiles and have unacceptable, unreliable fits. Their results should
not be used. These sources include T0526+24, T0556+19, 4C22.12,
T0629+10, 3C167. }

\end{deluxetable}

\clearpage

\begin{deluxetable}{ccc} 
\tablewidth{0pc} 

\tablecaption{Medians and Means of CNM $T_s$ \label{medianmeants} }

\tablehead{ \colhead{$b$-range} & \colhead{Median $T_s$} & 
\colhead{Mean $T_s$} }

\startdata
CNM, $|b|>10^\circ$, by $N_G$    & 48 & 88  \\
CNM, $|b|>10^\circ$, by $N(HI)$ & 70 & 108 \\
CNM, $|b|<10^\circ$, by $N_G$    & 47 & 71  \\
CNM, $|b|<10^\circ$, by $N(HI)$ & 63 & 99  \\
\enddata

\tablecomments{ Temperatures are in Kelvins. ``by $N_G$'' means that the
median and mean are taken over Gaussian components with no weighting by
$N(HI)$. ``by $N(HI)$'' means that half the column density lies above,
and half below, the median; and the mean is weighted by $N(HI)$. }

\tablecomments{ Figure \ref{histoplot4new} presents the histograms,
which have  long tails at high $T_s$ so that neither the median nor the
mean represent the typical values. }

\end{deluxetable}

\clearpage

\begin{deluxetable}{ccc} 
\tablewidth{0pc} 

\tablecaption{Medians and Means of $N(HI)$ \label{medianmeannh} }

\tablehead{ \colhead{$b$-range} & \colhead{Median $N(HI)_{20}$} & 
\colhead{Mean $N(HI)_{20}$}  }

\startdata
CNM, $|b|>10^\circ$   & 0.52 & 1.27  \\
CNM, $|b|<10^\circ$   & 1.97 & 5.00 \\
WNM, $|b|>10^\circ$   & 1.30 & 2.04  \\
WNM, $|b|<10^\circ$   & 8.13 & 12.03  \\
\enddata

\tablecomments{ $N(HI)_{20}$ is HI column density in units of $10^{20}$
cm$^{-2}$. Figure \ref{histoplotnhgaussians} presents the histograms.}

\end{deluxetable}

\clearpage

\include{tab4}

\clearpage

\include{tab5}

\clearpage

\begin{deluxetable}{ccccc} %\tablewidth{33pc}
\tablewidth{0pc} 

\tablecaption{GL's fluctuation statistics for two components in 3C348
 \label{glsourcepairs} }

\tablehead{ \colhead{$V_{LSR}$} & \colhead{$\tau_0$} & \colhead{$T_s$} &
\colhead{$N(HI)_{CNM}$} & \colhead{$\sigma(N(HI) \over N(HI)$}}

\startdata
0.5   &  $0.604 \pm 0.004$  & $32.5 \pm 5.8$ & 0.81 & 0.25 \\
--2.2 &  $0.259 \pm 0.003$  & $11.6 \pm 4.8$ & 0.10 & 0.10
\enddata

\tablecomments{ The first four columns are our Gaussian component data.
The fifth column is GL's rms profile area divided by the mean profile
area for the 5 positions listed in their Table 4.} 
 
\end{deluxetable}

\end{document}

%% file: tab4.tex
%\documentclass[preprint]{aastex}
%\usepackage{graphicx}
%\begin{document}

\begin{deluxetable}{ccccc}
%\tablewidth{33pc}
%\tabletypesize{\footnotesize}
\tabletypesize{\small}
\tablewidth{0pc}
\tablecaption{Spin versus H$_2$ Temperatures for proximate positions 
\label{h2tbl}}
\tablehead{
\colhead{Source}      & 
\colhead{$(l , b)$} &
\colhead{$N(HI)$} &
\colhead{$T$}
}

\startdata
\sidehead{NEAR $(l,b) = (160^\circ,-17^\circ)$:}
NRAO140  &  $(159.0, -18.8)$ & 13.4 & $ 27 \pm 13$ \\
3C93.1   &  $(160.0, -15.9)$ & 1.8  & $29 \pm 11 $ \\
HD21856  &  $(156, -17)$    & 11.0 & 84 \\
HD22951  & $159,-17) $       & 11.0 & 63 \\
HD23180  & $(160,-18)$       & 7.9  & 48 \\
%\vspace{4pt}
\sidehead{NEAR $(l,b) = (196^\circ,-13^\circ)$:}
HD24398  & $(162,-17)$       & 6.5  & 57 \\
3C142.1  & $(197.6,-14.5)$   & 7.0 & $49 \pm 16$ \\
HD36822  & $(195, -13)$      & 6.5      & 63  \\
%\vspace{4pt}
\sidehead{NEAR $(l,b) = (234^\circ,55^\circ)$:}
HD36861  & $( 195,-12 $      & 6.0  & 45 \\
3C245    & $(233.1, 56.3)$   & 0.5 & $510 \pm 8$ \\
HD91316  & $(235, 53)$       & 1.8 & 377
\enddata

\tablecomments{For radio source results, only the CNM component with the
largest $N(HI)$ is listed. $N(HI)$ is in units of $10^{20}$. Stars are
from Savage et al (1977). }

\end{deluxetable}

%\end{document}

%% file: tab5.tex
%\documentclass[preprint]{aastex}
%\usepackage{graphicx}
%\begin{document}

\begin{deluxetable}{cccrrrrrrr}
%\tablewidth{33pc}
\tabletypesize{\footnotesize}
%\tabletypesize{\small}
\tablewidth{0pc}
\tablecaption{CNM fluctuations for closely-spaced sources
\label{chtbl}}
\tablehead{
\colhead{Sources}      & 
\colhead{$ (l,b) $} &
\colhead{$ \Delta \theta $} &
\colhead{$ T_{exp} $} &
\colhead{$ \tau_0 $} &
\colhead{$ VLSR $} &
\colhead{$ FWHM $} &
\colhead{$ T_s $} &
\colhead{$ N(HI) $} &
\colhead{ AREA}
}

\startdata
\vspace{8pt}
${\rm 3C225a \over 3C225b}$  & $(220.0, 44.0)$ & 6.3 & ${5.8 \over 9.2}, 1.6$   & ${0.31 \over 0.75}, 2.5$   & ${4.0 \over 3.6}, 0.30 $   & ${1.3 \over 1.3}, 1.0$ & ${22 \over 17}, 1.2$ & ${0.17 \over 0.32}, 1.9$ & ${0.32 \over 0.73}, 2.3$  \\
\vspace{8pt}
${\rm 3C33-1 \over 3C33-2}$  & $(129.4,-49.3)$ & 4.2 & ${10.4 \over 10.2}, 1.0$ & ${0.034 \over 0.059}, 1.7$ & ${-4.6 \over -4.2}, 0.02 $ & ${9.4 \over 9.3}, 1.0$ & ${310 \over 178}, 1.7$   & ${2.0 \over 1.9}, 1.0$ & ${0.25 \over 0.43}, 1.7$  \\
\vspace{1pt}
${\rm 3C75-1 \over 3C75-2}$  & $(170.3,-44.9)$ & 3.4 & ${18 \over 17}, 1.1$ & ${0.73 \over 0.65}, 1.1$ & ${-10.4 \over -10.4}, 0.0 $ & ${2.1 \over 2.3}, 1.1$ & ${35 \over 36}, 1.0$   & ${1.0 \over 1.1}, 1.1$ & ${1.2 \over 1.2}, 1.0$  \\
\vspace{1pt}
${\rm 3C75-1 \over 3C75-2}$  & $(170.3,-44.9)$ & 3.4 & ${1.3 \over 3.1}, 2.3$ & ${0.082 \over 0.094}, 1.1$ & ${-6.1 \over -6.0}, 0.04 $ & ${3.0 \over 2.3}, 1.3$ & ${17 \over 34}, 2.1$   & ${0.08 \over 0.15}, 1.9$ & ${0.19 \over 0.18}, 1.0$  \\
\vspace{8pt}
${\rm 3C75-1 \over 3C75-2}$  & $(170.3,-44.9)$ & 3.4 & ${10 \over 10}, 1.0$ & ${0.13 \over 0.13}, 1.1$ & ${5.0 \over 4.9}, 0.02 $ & ${4.6 \over 4.4}, 1.0$ & ${84 \over 78}, 1.1$   & ${0.95 \over 0.93}, 1.0$ & ${0.45 \over 0.48}, 1.1$  \\
\vspace{1pt}
${\rm 3C98-1 \over 3C98-2}$  & $(179.8,-31.0)$ & 4.0 & ${1.7 \over 5.1}, 3.1$ & ${0.081 \over 0.090}, 1.1$ & ${-1.2 \over -1.5}, 0.09 $ & ${3.2 \over 3.2}, 1.0$ & ${21 \over 59}, 2.8$   & ${0.11 \over 0.33}, 3.0$ & ${0.21 \over 0.22}, 1.0$  \\
\vspace{1pt}
${\rm 3C98-1 \over 3C98-2}$  & $(179.8,-31.0)$ & 4.0 & ${7.8 \over 4.5}, 1.7$ & ${0.21 \over 0.20}, 1.0$ & ${9.4 \over 9.5}, 0.07 $ & ${1.5 \over 1.4}, 1.1$ & ${41 \over 24}, 2.8$   & ${0.25 \over 0.13}, 1.9$ & ${0.24 \over 0.21}, 1.1$  \\
\vspace{1pt}
${\rm 3C98-1 \over 3C98-2}$  & $(179.8,-31.0)$ & 4.0 & ${35 \over 36}, 1.0$ & ${0.37 \over 0.45}, 1.2$ & ${9.7 \over 9.6}, 0.07 $ & ${6.1 \over 4.6}, 1.3$ & ${115 \over 100}, 1.2$   & ${5.0 \over 4.1}, 1.2$ & ${1.4 \over 1.6}, 1.1$  \\
\vspace{8pt}
${\rm 3C98-1 \over 3C98-2}$  & $(179.8,-31.0)$ & 4.0 & ${5.9 \over 5.7}, 1.0$ & ${0.028 \over 0.035}, 1.3$ & ${22.8 \over 22.5}, 0.06 $ & ${5.4 \over 4.6}, 1.2$ & ${216 \over 166}, 1.3$   & ${0.63 \over 0.52}, 1.2$ & ${0.63 \over 0.52}, 1.2$  \\
\vspace{1pt}
${\rm 3C310 \over 3C315}$    & $(38.9,59.4)$   & 118 & ${18 \over 24}, 1.3$ & ${0.62 \over 0.78}, 1.3$ & ${-3.7 \over -4.2}, 0.25 $ & ${1.8 \over 2.2}, 1.2$ & ${39 \over 44}, 1.1$   & ${0.82 \over 1.5}, 1.8$ & ${0.85 \over 1.3}, 1.5$  \\
\vspace{1pt}
${\rm 3C310 \over 3C315}$    & $(38.9,59.4)$   & 118 & ${2.9 \over 8.3}, 2.9$ & ${0.061 \over 0.15}, 2.4$ & ${0.6 \over 1.6}, 0.21 $ & ${5.1 \over 4.4}, 1.2$ & ${49 \over 61}, 1.3$   & ${0.29 \over 0.77}, 2.7$ & ${0.24 \over 0.50}, 2.0$  \\
\enddata

\tablecomments{ Parameters for closely-spaced source pairs are listed as
fractions, with the numerators and denominators corresponding to the
appropriate component. Immediately to the right of each fraction we
write the ratio, which is always expressed as being $>1$ for purposes of
comparison among sources. For $VLSR$, however, instead of writing the
ratio we write the velocity difference divided by the linewidth $FWHM$. 
$\Delta \theta$ is the angular separation in arcmin, $T_{exp}$ the
central brightness of the component in the expected profile, $VLSR$ the
LSR velocity, $FWHM$ the line halfwidth, $T_s$ the spin temperature, 
$\tau$ the central opacity, $N(HI)$ the HI column density in units of
$10^{20}$ cm$^{-2}$, and AREA is 0.72 times the line area in km s$^{-1}$
(which is equal to $N(HI)_{20}$ if $T_s=40$ K).  }

\end{deluxetable}

%\end{document}